\documentclass[aps,pra,twocolumn,preprintnumbers,showpacs,superscriptaddress,amsmath,amssymb]{revtex4}
\usepackage{dcolumn}
\usepackage{bm}
\usepackage{amssymb}
\usepackage[german, english]{babel}
\usepackage{graphicx}
\usepackage{color}
\usepackage[letterpaper,total={7in,9.5in},top=0.75in,left=0.75in]{geometry}

\usepackage{array}

\begin{document}
\title{Reservoir spectroscopy of 5s5p\,{$^3$P$_2$} -- 5s\textit{n}d\,{$^3$D$_{1,2,3}$} transitions in strontium}

\author{Simon Stellmer}
\affiliation{Institut f\"ur Quantenoptik und Quanteninformation (IQOQI), \"Osterreichische Akademie der Wissenschaften, 6020 Innsbruck, Austria}
\affiliation{Institut f\"ur Experimentalphysik und Zentrum f\"ur Quantenphysik, Universit\"at Innsbruck, 6020 Innsbruck, Austria}
\affiliation{Vienna Center for Quantum Science and Technology, Atominstitut, TU Wien, 1020 Vienna, Austria}
\author{Florian Schreck}
\affiliation{Institut f\"ur Quantenoptik und Quanteninformation (IQOQI),
\"Osterreichische Akademie der Wissenschaften, 6020 Innsbruck, Austria}
\affiliation{Institute of Physics, University of Amsterdam, 1098 XH Amsterdam, The Netherlands}

\date{\today}

% 32.20.Jc Atomic spectra, visible and ultraviolet spectra
% 37.10.De Atom cooling methods

\begin{abstract}
We perform spectroscopy on the optical dipole transitions 5s5p\,{$^3$P$_2$} -- 5s$n$d\,{$^3$D$_{1,2,3}$}, $n \in (5,6)$, for all stable isotopes of atomic strontium. We develop a new spectroscopy scheme, in which atoms in the metastable $^3$P$_2$ state are stored in a reservoir before being probed. The method presented here increases the attained precision and accuracy by two orders of magnitude compared to similar experiments performed in a magneto-optical trap or discharge. We show how the state distribution and velocity spread of atoms in the reservoir can be tailored to increase the spectroscopy performance. The absolute transition frequencies are measured with an accuracy of 2\,MHz. The isotope shifts are given to within 200\,kHz. We calculate the $A$ and $Q$ parameters for the hyperfine structure of the fermionic isotope at the MHz-level. Furthermore, we investigate the branching ratios of the $^3$D$_{J}$ states into the $^3$P$_{J}$ states and discuss immediate implications on schemes of optical pumping and fluorescence detection.
\end{abstract}

\maketitle

\section{Introduction}

The rich electronic structure of elements with two valence electrons, comprising long-lived metastable states and ultranarrow transitions, is of great interest for precision measurements, as demonstrated by optical clocks \cite{Ido2003rfs,Takamoto2005aol,Derevianko2011poo,Hinkley2013aac,Bloom2014aol} or gravimeters \cite{Poli2011pmo}. Quantum degenerate samples of Yb \cite{Takasu2003ssb,Fukuhara2007bec,Fukuhara2007dfg,Fukuhara2009aof,Sugawa2011bec} and the alkaline-earth elements Ca \cite{Kraft2009bec,Halder2012art} and Sr \cite{Stellmer2009bec,MartinezdeEscobar2009bec,Mickelson2010bec,DeSalvo2010dfg,Tey2010ddb,Stellmer2010bec,Stellmer2013poq} have opened new possibilities, such as the study of SU($N$) magnetism \cite{Cazalilla2014ufg,Wu2003ess,Wu2006hsa,Cazalilla2009ugo,Hermele2009mio,Gorshkov2010tos,FossFeig2010ptk,Xu2010lim,Hung2011qmi,Taie2010roa,Taie2012asm,Martin2013aqm,Zhang2014soo,Scazza2014oot}, novel schemes to simulate gauge fields \cite{Gerbier2009gff,Cooper2011ofl,Beri2011zti,Gorecka2011smf,Dalibard2011agp,Banerjee2013aqs}, the engineering of interactions beyond contact interactions \cite{Diehl2010did,Bhongale2013qpo,Olmos2013lri,Lahrz2014dqi}, the creation of driven-dissipative many-body states \cite{Yi2012ddm}, the simulation of an extra dimension \cite{Boada2012qso}, and new ways to perform quantum computation \cite{Stock2008eog,Daley2008qcw,Gorshkov2009aem,Reichenbach2009cns,Daley2011qca}.

Many of these applications and proposals rely on broad optical transitions originating from the metastable $^3$P$_{0,1,2}$ triplet states. These transitions are essential for laser cooling of Mg, Ca, and Sr \cite{Rehbein2007oqo,Binnewies2001dca,Katori1999mot}. In quantum computation schemes, they can be employed to drive Raman transitions between the metastable states \cite{Daley2008qcw}. Transitions originating from the $^3$P$_2$ state are particularly important:~A suitable $^3$P$_2$ -- {$^3$D$_3$} transition could be used for operation of a magneto-optical trap (MOT) \cite{Grunert2002sdm,Feldker2011mot}, for fluorescence detection of single atoms, and for absorption imaging. Precise knowledge of these transitions is also required for the calculation of the polarizability of the $^3$P$_2$ state, which is important for quantum computation schemes \cite{Porsev2008dos,Daley2008qcw}. The frequencies and strengths of the transitions originating from the metastable states are needed to calculate the black-body radiation (BBR) shifts of a clock transition, which is particularly important for strontium \cite{Porsev2008dos,Safranova2013brs}. Atoms in the $^3$P$_2$ state can be used to study systems with quadrupolar interactions \cite{Bhongale2013qpo,Lahrz2014dqi}. Furthermore, magnetic Feshbach resonances have been observed between atoms in the $^3$P$_2$ and $^1$S$_0$ states \cite{Kato2013cor}.

\begin{figure}[b]
\includegraphics[width=\columnwidth]{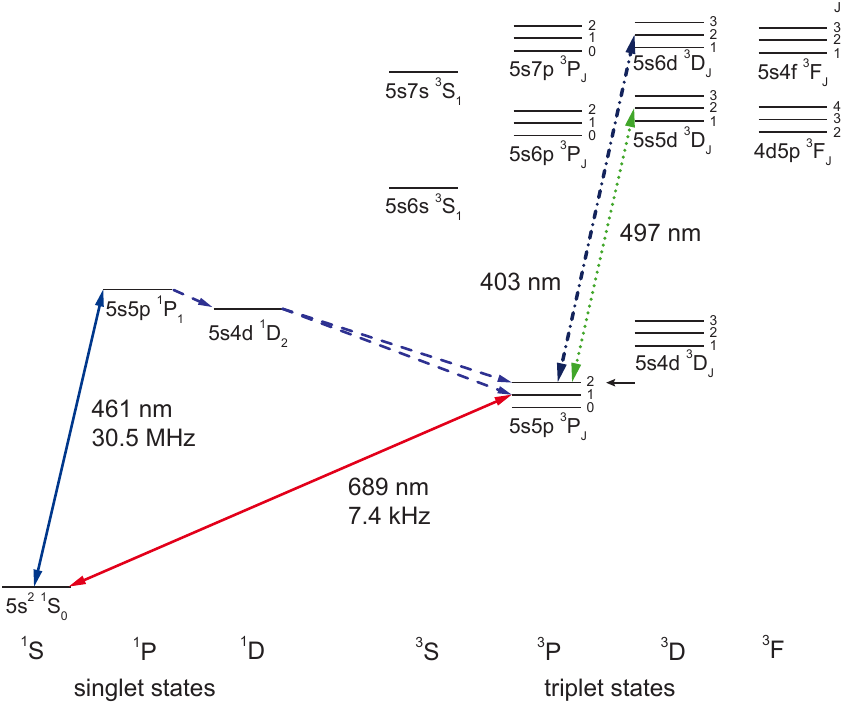}
\caption{Schematic illustration of the energy levels and transitions in strontium relevant for this work. Atoms in the metastable 5s5p\,{$^3$P$_2$} state (indicated by the small solid arrow) are repumped using the 5s5p\,{$^3$P$_2$} - 5s$n$d\,{$^3$D$_{1,2,3}$} transitions around 497\,nm ($n=5$, dotted line) and around 403\,nm ($n=6$, dash-dotted line).}
\label{fig:fig1}
\end{figure}

So far, spectroscopic data is available for only some of the dipole transitions within the triplet system of strontium; see Ref.~\cite{Sansonetti2010wtp} for a recent compilation. The 5s5p\, $^3$P$_0$ -- 5s6s\,{$^3$S$_1$} and 5s5p\,{$^3$P$_1$} -- 5s6s\,{$^3$S$_1$} transitions, as well as the isotope shifts and hyperfine splittings involved, have been measured to a precision of a few 100\,kHz \cite{Courtillot2005aso}. The Rydberg series 5s$n$d\,{$^3$D$_{1,2,3}$} for $n>12$ has been investigated in a discharge \cite{Beigang1982opl}, and the series 5s$n$d\,{$^3$D$_2$} for $n>4$ has been studied by multi-photon ionization \cite{Esherick1977bep}. The 5s5p\,{$^3$P$_2$} -- 5s4d\,{$^3$D$_2$} transition has been investigated using atoms trapped in a MOT \cite{Mickelson2009ras}; these three experiments reached an accuracy of about 1\,GHz and a precision of a few 10\,MHz.

In this article, we present a new spectroscopy scheme to probe transitions originating from the 5s5p\,{$^3$P$_2$} state in strontium. Spectroscopy is performed for the three stable bosonic isotopes $^{84, 86, 88}$Sr as well as the fermionic isotope $^{87}$Sr. While the bosonic isotopes have zero nuclear spin, the fermionic one has a nuclear spin $I=9/2$, which gives rise to a rich hyperfine structure. The wealth of hyperfine transitions requires a sophisticated spectroscopy scheme. Specifically, we use a magnetic trap as a reservoir to accumulate atoms in the metastable $^3$P$_2$ state, hence we name this method \emph{reservoir spectroscopy}. Atoms in this trap are then probed, and we can populate or deplete certain hyperfine states beforehand to improve measurement performance. This scheme is used to measure the ``green'' 5s5p\,{$^3$P$_2$} -- 5s5d\,{$^3$D$_{1,2,3}$} transitions at 497\,nm and the ``blue'' 5s5p\,{$^3$P$_2$} -- 5s6d\,{$^3$D$_{1,2,3}$} transitions at 403\,nm, all of which have not yet been investigated spectroscopically. We present two different approaches to determine the number of repumped atoms, namely fluorescence detection and recapture into a narrowline MOT, followed by absorption imaging. Both precision and accuracy are improved by about two orders of magnitude compared to similar spectroscopy measurements in a MOT or discharge. Note that related schemes have recently been used for spectroscopy in Ca \cite{Dammalapati2011lso} and to determine the lifetimes of the 3s3p\,{$^3$P$_2$} state in Mg \cite{Jensen2011edo} and the 5s6d\,{$^3$D$_1$} state in Yb \cite{Beloy2012dot}. Furthermore, we calculate the branching ratios of all relevant $^3$D$_J$ $\rightarrow$ {$^3$P$_J$} decay channels, compare our findings to experimentally determined values, and comment on possible implications for future experiments.

This article is structured as follows: In Sec.~\ref{sec:review}, we give a short review over various repumping transitions in strontium. We then describe the spectroscopy scheme used in the present work (Sec.~\ref{sec:scheme}). The spectroscopy data is given in Sec.~\ref{sec:data}. The calculation of relevant branching ratios is presented together with experimental investigations of the repump efficiency (Sec.~\ref{sec:efficiency}) and the imperfection of quasi-cycling $^3$P$_2$ -- $^3$D$_3$ transitions (Sec.~\ref{sec:cycling}). A short conclusion follows in Sec.~\ref{sec:conclusion}, a detailed error analysis is given in the Appendix.

\section{Repumping in strontium}
\label{sec:review}

The level structure of strontium and the other alkaline-earth elements naturally suggests the broad $^1$S$_0$ -- $^1$P$_1$ dipole transition for the first laser cooling stage. Cooling on this transition is simplified through the absence of hyperfine structure in the $^1$S$_0$ ground state, such that repumping from hyperfine states, very common to alkali MOTs, is not required. There are, however, long-lived electronic states with energies smaller than the $^1$P$_1$ state. The performance of a MOT operated on the  $^1$S$_0$ -- $^1$P$_1$ transition may therefore be reduced, as an excited atom can decay into one of these states and leave the MOT cycle. This is reminiscent of the cases of Er \cite{McClelland2006lcw} and Dy \cite{Lu2010tud}, but the intermediate states in these cases are short-lived.

There are various decay pathways from the  $^1$P$_1$ state into the long-lived $^3$P$_{0,1,2}$ states; see Fig.~\ref{fig:fig1}. The dominant decay channel is opened up by the 5s4d\,{$^1$D$_2$} state which is located below the 5s5p\,{$^1$P$_1$} state. The branching ratio from the $^1$P$_1$ state into the $^1$D$_2$ stage is roughly $1:50\,000$ for Ca and Sr, and roughly $1:300$ for Ba and Ra. The atoms further decay into the $^3$P$_{1,2}$ metastable triplet states with a branching ratio of $2:1$. The second decay channel is of the type $^1$P$_1$ $\rightarrow$ $^3$D$_{1,2}$ $\rightarrow$ $^3$P$_{0,1,2}$ and is about two orders of magnitude weaker:~for the first step, a branching ratio of $1:3.6 \times 10^6$ is given in Ref.~\cite{Bidel2002per}, and we estimate that about one third of the atoms end up in the $^3$P$_0$ state. This type of loss channel is the dominant one in $^1$S$_0$ -- $^1$P$_1$ MOTs of Yb \cite{Porsev1999eda,Cho2012oro}. The transition probabilities of a third decay channel, the direct decay  $^1$P$_1$ $\rightarrow$ $^3$P$_{0,1,2}$, have not yet been calculated, but we expect them to be even smaller. We conclude that all three $^3$P$_{0,1,2}$ states are eventually populated during operation of the MOT, where the relative rates are about $<0.01 : 2 : 1$, respectively.

The three metastable $^3$P$_J$ states have very different lifetimes. In Sr, the $^3$P$_1$ state has a comparably short lifetime of $21\,\mu$s, and atoms arriving in this state decay back into the $^1$S$_0$ state and into the MOT cycle well before leaving the MOT region. The $^3$P$_0$ and $^3$P$_2$ states, however, have much longer lifetimes. The lifetime of the $^3$P$_0$ state has been calculated to be thousands of years for bosonic atoms, and around 100\,s for fermionic atoms \cite{Santra2004pom}. Atoms in this state are lost from the MOT region. The lifetime of the $^3$P$_2$ state has been measured to be around 500\,s in absence of BBR \cite{Yasuda2004lmo}. The BBR field of the environment, however, can drive transitions $^3$P$_2$ $\rightarrow$ $^3$D$_J$, from where the atoms decay into the $^3$P$_{0,1}$ states. This process reduces the lifetime of the $^3$P$_2$ state to a few 10\,s at room temperature \cite{Xu2003cat,Yasuda2004lmo}, which we verify in our experiment. Atoms in the $^3$P$_2$ state can naturally be trapped in the quadrupole field of the MOT, provided that they are in a low-field seeking $m_J$ state and have a kinetic energy smaller than the trap depth \cite{Katori2001lco}.

Various transitions can be employed for repumping, and the choice of the transition is guided by both the anticipated repump efficiency and the availability of suitable lasers at the respective wavelengths. A number of transitions have already been used for repumping. An early experiment tried to close the leakage of atoms into the triplet states by pumping them directly from the $^1$D$_2$ state into the 5s6p\,{$^1$P$_1$} state at 717\,nm \cite{Kurosu1992lca}. This approach is inefficient due to a significant branching ratio from the 5s6p\,{$^1$P$_1$} state into the triplet states. Other experiments use a repump laser at 707\,nm to pump atoms from the $^3$P$_2$ state via the $^3$S$_1$ state into the $^3$P$_1$ state, from where they rapidly decay into the $^1$S$_0$ ground state \cite{Dinneen1999cco}. A large fraction of atoms in the intermediate $^3$S$_1$ state decays into the $^3$P$_0$ state, necessitating a second laser at 679\,nm to repump this state as well. This repumping approach rigorously collects atoms from all possible decay paths and facilitates MOT lifetimes of up to ten seconds. A third strategy involves any of the 5s$n$d\,{$^3$D$_2$} states at $3.01\,\mu$m \cite{Mickelson2009ras}, 497\,nm \cite{Poli2005cat}, or 403\,nm for $n=4,5,6$, respectively. This strategy requires only one laser. Repumping via the 5p$^2$\,{$^3$P$_2$} state at 481\,nm is also efficient \cite{KillianPrivComm}, as favorable branching ratios limit the undesired decay into the $^3$P$_0$ state to below 0.5\%. When repumping only the $^3$P$_2$ state via a $^3$D$_2$ state, loss through the $^3$P$_0$ state persists and limits the lifetime of a continuously repumped MOT to about one second. Loss into the $^3$P$_0$ state also originates from leakages in the repump process; see Sec.~\ref{sec:efficiency}.

Atoms in the metastable states can be returned to the ground state either during or after the MOT stage. Generally, continuous repumping allows for a faster accumulation of atoms, as most of the atoms falling into the $^3$P$_2$  state appear in non-trapped $m_F$ states and are lost if not quickly repumped. Repumping of the $^3$P$_0$ state improves the MOT atom number and lifetime even further. The MOT atom number is determined by the loading rate, which is about $10^9$ $^{88}$Sr atoms/s for our experiment, and limited by excited-state collisions, where the loss rate coefficient has been determined to $\beta=4.5\times 10^{-10}$\,cm$^3$/s \cite{Dinneen1999cco}. This strategy of continuous repumping is followed in optical clock experiments.

A second scheme accumulates atoms in the metastable reservoir and transfers them into the ground state after the MOT has been extinguished \cite{Katori2001lco,Stuhler2001clo}. This accumulation strategy is advantageous when large atom numbers are required. The density limitation of the reservoir arises from inelastic two-body collisions, where the loss rate coefficient has been quantified to be $\beta\approx 1 \times 10^{-10}$\,cm$^3$/s \cite{Traverso2009iae}, but the large volume of the reservoir ($\sim 20\,$cm$^3$) ensures that two-body collisions are rare. After saturation, the reservoir will contain roughly 1000-times more atoms than the MOT, where the factor is given by the ratio of the lifetime of the $^3$P$_2$ state (roughly 20\,s) and the lifetime of the MOT (roughly 20\,ms). In our experiment, we can load the reservoir with up to $10^{10}$ atoms of $^{88}$Sr. Accumulation in the reservoir also allows to sequentially load multiple isotopes \cite{Poli2005cat}. Experiments with the fermionic $^{87}$Sr isotope will benefit from the storage in the $^3$P$_2$ state, as the hyperfine structure impedes both MOT operation and repumping. This strategy is also preferred when repumping of the $^3$P$_0$ state is not available \cite{Stellmer2009bec,MartinezdeEscobar2009bec}. 

The repump transitions discussed above are dipole-allowed and have typical linewidths of a few MHz and saturation intensities of a few mW/cm$^2$. The isotope shifts are at most 100\,MHz. In the case of $^{87}$Sr, efficient repumping is complicated by the hyperfine structure of the states involved. We find that all five hyperfine states $F=5/2$ through $F=13/2$ of the $^3$P$_2$ level are populated during the MOT phase, however at different relative amounts:~roughly 80\% of the atoms populate the $F=13/2$ and $F=11/2$ states. The hyperfine splittings of the $^3$P$_2$, $^3$S$_1$, and $^3$D$_2$ states are on the order of a few GHz. In a typical experimental cycle, repumping is performed only on the $F=11/2 \rightarrow F'=13/2$ and $F=13/2 \rightarrow F'=13/2$ transitions, or the laser is rapidly scanned across all hyperfine transitions.

In our experiment, the repump light illuminates the entire MOT region with an intensity of typically $0.05\,I_{\mathrm{sat}}$, corresponding to a few 100\,$\mu$W. Here, $I_{\rm sat}=\pi h c A_{ik} / 3 \lambda^3$ is the saturation intensity of the transition, where $A_{ik}$ is the transition probability betweens states $i$ and $k$, and $\lambda$ is the wavelength of the transition. Repumping of the reservoir typically takes a few milliseconds.

\section{Spectroscopy scheme}
\label{sec:scheme}

In this section, we will describe how the reservoir of magnetically trapped atoms can be used to improve spectroscopy performance by orders of magnitude compared to other approaches. We will present two complementary detection schemes and characterize our approach.

\subsection{Reservoir spectroscopy}
\label{sec:reservoir_spectroscopy}

A straightforward approach to determine the frequency of a repump transition is to monitor the MOT fluorescence while scanning the repump laser frequency \cite{Mickelson2009ras}. This works well if the repumping increases the MOT atom number $N_{\rm MOT}$ significantly. Typical values of $(N_{\rm rMOT}-N_{\rm MOT})/N_{\rm MOT}$, where $N_{\rm rMOT}$ is the MOT atom number in presence of the repump light, range between 30 for a non-saturated bosonic MOT and 0.1 for repumping only one hyperfine state of the $^{87}$Sr isotope, to below 0.001 for transitions with poor repump efficiency. Considering that the MOT atom number is typically stable only to within 1\%, a more sophisticated scheme is needed for weak repump transitions.

\begin{figure}
\includegraphics[width=\columnwidth]{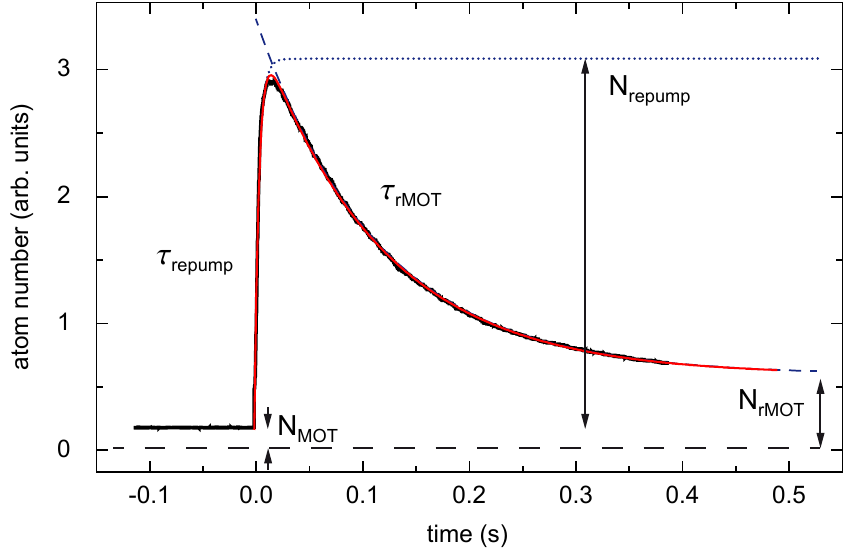}
\caption{Burst in MOT fluorescence after repumping of atoms in the $^3$P$_2$ reservoir state back into the MOT cycle. The data is approximated by the sum of two exponential functions (solid red line), consisting of a fast growing ($\tau_{\rm repump}$, dotted blue line) and a slowly decreasing contribution ($\tau_{\rm rMOT}$, dashed blue line). The horizontal dashed line indicates the small and stable background signal, caused by residual stray light.}
\label{fig:fig2}
\end{figure}

A scheme that employs atoms in the reservoir has the potential to drastically improve the signal. The number of atoms in the reservoir, $N_{\rm res}$, can be 1000 times larger than $N_{\rm MOT}$, providing a tremendous leverage to spectroscopically resolve transitions with very poor repump efficiency. Illuminating a fully charged reservoir with repump light in presence of the MOT light induces a burst in fluorescence \cite{Katori2001lco,Dammalapati2011lso} that can exceed the fluorescence of the bare MOT by a factor of 1000; see Fig~\ref{fig:fig2}. 

After switching the repump light on at time $t_0=t(0\,{\rm s})$, the reservoir atoms are quickly pumped back into the MOT cycle on a timescale $\tau_{\rm repump}$, which is a few milliseconds, leading to a burst in fluorescence. Subsequently the MOT atom number slowly decreases due to the decay of atoms into the $^3P_0$ state, which is not repumped in our experiments, and possibly other loss processes \cite{Dinneen1999cco}. This timescale $\tau_{\rm rMOT}$ ranges between 10\,ms and 1\,s, depending on the efficiency of the repump transition used.

We use a simple model of the form
\begin{equation*}
N(t) =  N_{\rm MOT} + N_{\rm repump} \times (e^{-\frac{t-t_0}{\tau_{\rm rMOT}}} - e^{-\frac{t-t_0}{\tau_{\rm repump}}})
\end{equation*}
to describe the evolution of the MOT atom number for times $t_0<t<t_0 + \tau_{\rm rMOT}$. Here, we assume $\tau_{\rm repump}$ to be much smaller than $\tau_{\rm rMOT}$. Fitting this approximation to our data allows us to extract all relevant parameters. The number of repumped atoms, $N_{\rm repump}$, certainly depends on the frequency of the repump light, and ideally approaches $N_{\rm res}$ near the resonance position. We find that $N_{\rm repump}/N_{\rm MOT}$ can assume values of up to 1000, clearly showing the advantage of repump spectroscopy.

The entire experimental sequence used for reservoir spectroscopy is as follows. Atoms emerging from an effusive source are Zeeman-slowed and cooled in a MOT using the broad $^1$S$_0$ -- $^1$P$_1$ transition at 461\,nm. The lifetime of the MOT is about 20\,ms and depends on the intensity of the MOT beams, which sums to about $0.1\,I_{\rm sat}$. Atoms accumulate in the low-field seeking $m_J=+1$ and $m_J=+2$ components of the long-lived $^3$P$_2$ state, and we operate the MOT until the reservoir is charged with about $10^7$ atoms. This takes between 50\,ms and 10\,s, depending on the natural abundance of the respective isotope. Afterwards, we turn on the repump light, which illuminates the entire reservoir volume at an intensity of $10^{-3}\,I_{\rm sat}$ and returns the atoms back into the ground state within a few milliseconds. While the $^3$P$_2$ -- $^3$D$_2$ transition is commonly used for repumping, also the $^3$P$_2$ -- $^3$D$_{1,3}$ transitions return atoms into the ground state, however with lower efficiency. Further details of the apparatus used for this experiment can be found in Ref.~\cite{Stellmer2013poq}.

\subsection{Fluorescence detection}
\label{sec:fluorescence}

The MOT fluorescence is measured by imaging the MOT onto a large-area photo diode with a magnification of 0.5, capturing 0.6\% of the total fluorescence. For low densities of the MOT, the fluorescence is roughly proportional to the atom number. We vary the frequency of the repump light in consecutive experimental runs and determine $N_{\rm repump}$ for each run. These values are then fitted by a Lorentzian lineshape and allow us to determine the resonance position to at best 300\,kHz, where typical uncertainties are about 1\,MHz. This scheme of fluorescence detection is very robust and has previously been used to determine the lifetime of the $^3$P$_2$ state \cite{Yasuda2004lmo}. Note that, for small values of $N_{\rm repump}$, it is advantageous to remove the background signal $N_{\rm MOT}$ by removing all MOT atoms prior to the flash of repump light. This can easily be done by blocking the atomic beam or extinguishing the Zeeman slower light prior to repumping.

\subsection{Absorption imaging}
\label{sec:absorption}

We developed a more sophisticated detection method to overcome certain drawbacks of fluorescence imaging. The amplitude of the fluorescence signal depends directly on the intensity and polarization of the MOT beams, which might fluctuate and add noise to the signal. These issues eventually limit the performance and led us to develop a scheme based on absorption imaging. After loading the reservoir with $10^7$ atoms, we extinguish the MOT light. A mechanical shutter blocks the atomic beam, and a 1-s wait ensures that only atoms in the magnetic trap remain in the probe volume. Upon a short repump flash, the atoms are captured in a narrow-line MOT operated on the $^1$S$_0$ -- $^3$P$_1$ intercombination line \cite{Katori1999mot,Mukaiyama2003rll}. The MOT light is initially frequency-broadened to a few MHz to increase the capture efficiency. Within 500\,ms, the frequency broadening is removed, and the intensity is lowered to $1\,I_{\rm sat}$. The atoms are cooled to a temperature below $1\,\mu$K and compressed to a dense cloud with a size of order $100\,\mu$m. This stage increases the phase-space sensity by 8 orders of magnitude compared to the broad-linewidth $^1$S$_0$ -- $^1$P$_1$ MOT.

Recapture into the narrow-line MOT rather than the broad-line MOT has numerous advantages:~the narrow-line MOT is perfectly cycling, excited-state collisions are rare, and the small size and low temperature are ideal for subsequent absorption imaging. Imaging is performed on the $^1$S$_0$ -- $^1$P$_1$ transition and allows for a background-free determination of the atom number. This technique allows us to reduce the uncertainty in the fit of the Lorentzian profile to below 10\,kHz for typical settings.

\subsection{Characterization}
\label{sec:characterization}

\begin{figure*}
\includegraphics[width=179mm]{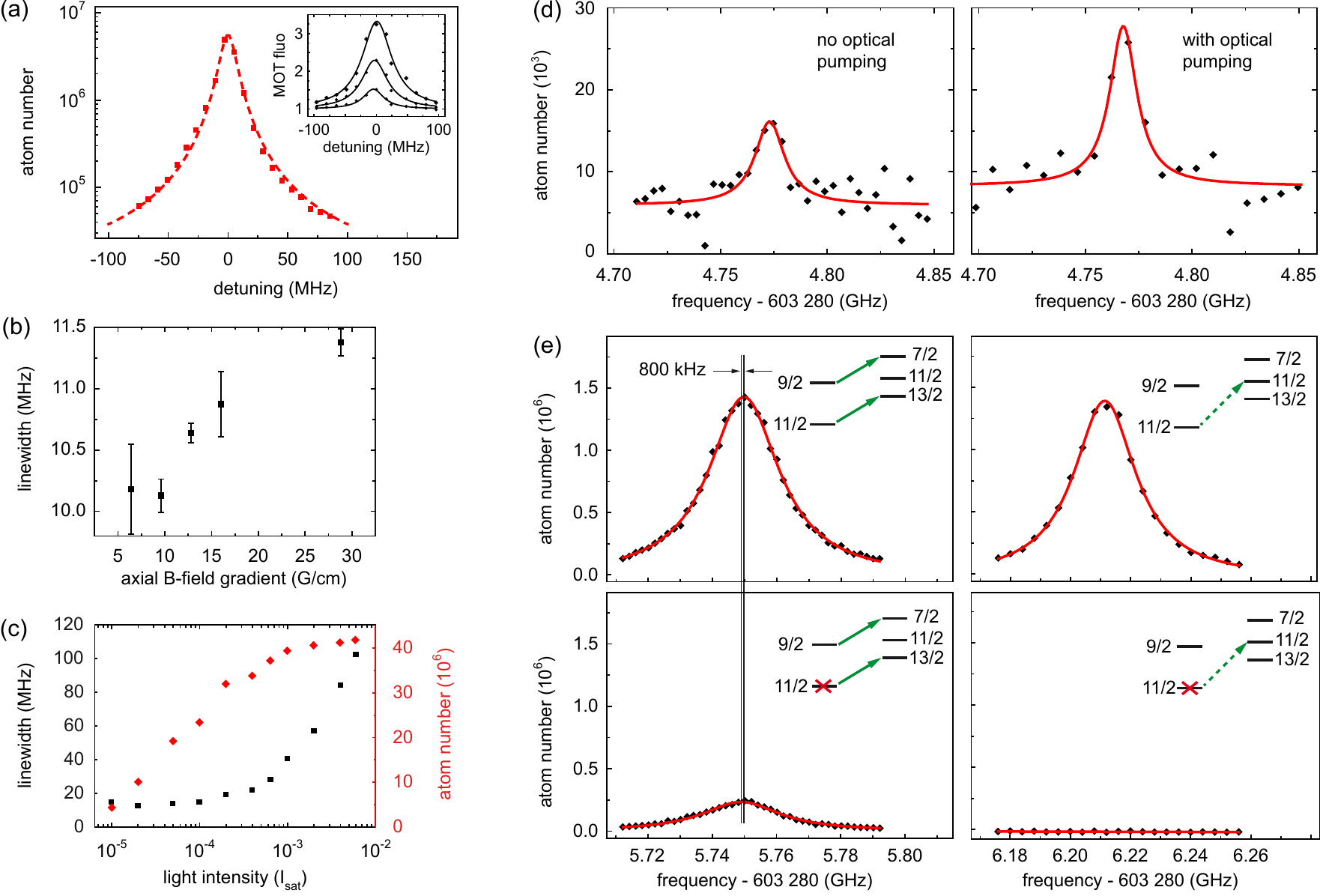}
\caption{Reservoir spectroscopy scheme. (a) The reservoir spectroscopy signal is background-free and rises many orders of magnitude above the background noise. By contrast, the fluorescence signal of a MOT (inset) is increased by only a small factor on resonance. A Lorentzian profile is fit to each data set, and the fluorescence curves are taken at a repump laser intensity of $I=1$, 3, and $10\times 10^{-3}\,I_{\mathrm{sat}}$, from below. (b) Dependence of the width of the transition on the depth of the magnetic trap. (c) Width (black squares) and atom number (red circles) in dependence of repump intensity. (d) Accumulation of a certain hyperfine state by optical pumping increases the signal. (e) Depletion of a certain hyperfine state allows for a separation of otherwise overlapping hyperfine transitions. Measurements (a) through (c) are taken on the 5s5p\,{$^3$P$_2$} -- 5s5d\,{$^3$D$_2$} transition of $^{88}\mathrm{Sr}$, measurements (d) and (e) employ $^{87}\mathrm{Sr}$; see the text for details.}
\label{fig:fig3}
\end{figure*}

We will now compare the reservoir spectroscopy scheme with the approach of monitoring the increase of the broad-line MOT fluorescence upon repumping \cite{Mickelson2009ras}.We will also   introduce two methods to increase the signal strength. The detection scheme considered in the following is the absorption imaging scheme, but the conclusions also hold for fluorescence detection after recapture into the broad MOT. 

Reservoir spectroscopy is advantageous to direct spectroscopy in the MOT in three ways. First, our measurement is intrinsically background-free. Far away from the repump resonance, no atoms are captured in the narrow-line MOT. The small background noise on the absorption images stems from photon shot noise, which leads to noise in the atom number corresponding to at most 1000 atoms. This background noise can be reduced by averaging over multiple images. On the other hand, spectroscopy in a MOT will always incoperate the background of the MOT itself. The superior signal strength is shown in Fig.~\ref{fig:fig3}(a). Depending on the MOT loading rate and repump intensity used, the fluorescence of the MOT increases by a factor of up to 30, exhibiting a substantially broadened linewidth. The atom number increase for the fermionic isotope is at most a few 10\% on resonance due to the involved hyperfine structure. The reservoir spectroscopy described here can yield a signal without intrinsic background and an amplitude more than three orders of magnitude larger than the background noise. The inset of Fig.~\ref{fig:fig3}(a) shows typical fluorescence spectra of a $^{88}\mathrm{Sr}$ MOT at repump intensities of about $1$, $3$, and $10\times10^{-3}\, I_{\mathrm{sat}}$, starting from below (solid black lines, normalized to the fluorescence of a MOT without repumping). The dashed red curve shows the corresponding measurement using reservoir spectroscopy, taken at an intensity of $5\times10^{-5}\, I_{\mathrm{sat}}$. The amplitude of $5\times 10^6$ atoms is more than three orders of magnitude above the background noise, and the width is smaller by a factor of four compared to the fluorescence curve at highest repump intensity. 

The second advantage of reservoir spectroscopy is the potential to reduce the Doppler broadening:~the quadrupole field, which confines the metastable atoms, acts as a velocity filter, as it traps only atoms below a certain kinetic energy. We typically set the axial gradient of the quadrupole field to 55\,G/cm during the MOT phase. Atoms may roam the magnetic trap up to an axial radius of 11\,mm, given by geometric constraints of our vacuum cell. For bosonic atoms in the $^3$P$_2$,\,$m_J=+1$ state, the trap depth is $U=4\,$mK, which is slightly larger than the Doppler temperature of $T_D=0.72\,$mK. The maximally trappable velocity generally depends on the $m_F$ state, and, for the fermionic isotope, also on the hyperfine state due to differing $g_F$ factors. A deeper trap will thus capture more atoms also at higher velocity classes, while the confinement can be reduced to deliberately select only the low-velocity atoms to reduce the Doppler-broadening, as shown in Fig.~\ref{fig:fig3}(b). In this measurement, the quadrupole field is ramped down after the MOT phase to adiabatically cool the sample and to allow high-energy atoms to escape.

As a third advantage, we can selectively populate the reservoir with the specific states of interest. This is important for the fermionic isotope with its large number of hyperfine levels. In particular, atoms can be pumped into the state of interest during or after the MOT phase. This is illustrated in Fig.~\ref{fig:fig3}(d): During the $^{87}\mathrm{Sr}$ MOT, atoms in all five $^3$P$_2$ hyperfine states are accumulated in the reservoir, however at different relative amounts. The $F=7/2$ state has the smallest $g_F$-factor, therefore only very few atoms are trapped. In addition, $\pi$-transitions into the $^3$D$_2$ states tend to be weaker than $\sigma$-transitions, making the $F=7/2 \rightarrow F'=7/2$ transition one of the weakest among all 13 $^3$P$_2$ -- $^3$D$_2$ transitions. Constantly pumping atoms from the $F=9/2$ into the $F=7/2$ state by standard optical pumping can increase the population of this state significantly.

Similarly, atoms from undesired states can be removed from the reservoir. This is shown in Fig.~\ref{fig:fig3}(e): The $F=11/2 \rightarrow F'=13/2$ and $F=9/2 \rightarrow F'=7/2$ hyperfine transitions of the 5s5p\,{$^3$P$_2$} -- 5s5d\,{$^3$D$_2$} line in $^{87}$Sr overlap entirely. The initial population of the $F=11/2$ state is about ten times that of the $F=9/2$ state, therefore the $F=11/2 \rightarrow F'=13/2$ transition completely covers the $F=9/2 \rightarrow F'=7/2$ transition. To selectively interrogate only the $F=9/2 \rightarrow F'=7/2$ transition, we remove all $F=11/2$-state atoms during the MOT phase by optical pumping into the $F'=11/2$ state. As an additional benefit, this also increases the $F=9/2$ population. We first verify on the $F=11/2 \rightarrow F'=11/2$ (or alternatively the $F=11/2 \rightarrow F'=13/2$) transition that all $F=11/2$ atoms have been removed, and then selectively probe the $F=9/2$ state and determine the frequency difference between the $F=11/2 \rightarrow F'=13/2$ and $F=9/2 \rightarrow F'=7/2$ transitions to be $800(200)\,$kHz. This would not be possible by spectroscopy on a continuously operated MOT.

To optimize our spectroscopy approach, we measure the linewidth of the transition and the amplitude of the signal in dependence of repump intensity. A larger repump intensity will yield a larger amplitude of the signal, but also a larger width; see Fig.~\ref{fig:fig3}(c). We find an intensity region around $5\times 10^{-4}\, I_{\mathrm{sat}}$ where the width is not broadened by the intensity, but the amplitude has almost saturated. In this region, width and amplitude are almost independent of the repump time for times greater than about 10\,ms. A distortion of the line shape can be observed only for intensities lower than $2\times 10^{-4}\, I_{\mathrm{sat}}$. For higher intensities, we do not observe any intensity-dependent line shifts.

\section{Spectroscopic data}
\label{sec:data}

\begin{figure}
\includegraphics[width=\columnwidth]{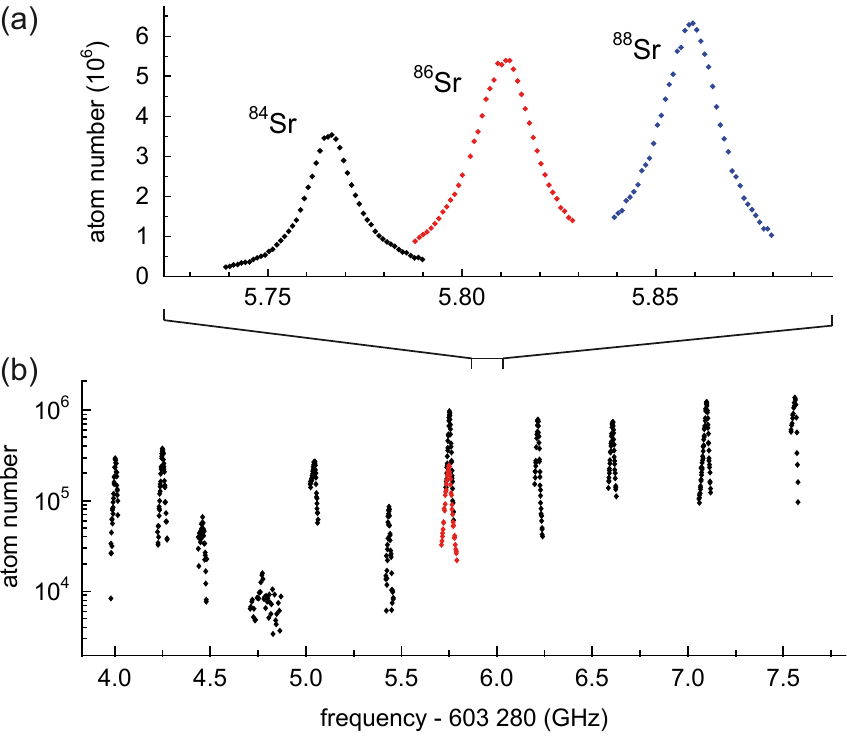}
\caption{Reservoir spectroscopy data obtained with the absorption imaging detection method, shown here for the 5s5p\,{$^3$P$_2$} -- 5s5d\,{$^3$D$_2$} transition. (a) Transitions of the bosonic isotopes $^{84}\mathrm{Sr}$ (black, left peak), $^{86}\mathrm{Sr}$ (red, center), and $^{88}\mathrm{Sr}$ (blue, right). (b) A scan across the transitions of the fermionic $^{87}\mathrm{Sr}$ isotope. The separate measurement of the two overlapping lines around 5.75\,GHz is discussed in Sec.~\ref{sec:characterization}.}
\label{fig:fig4}
\end{figure}

\subsection{5s5p\,{$^3$P$_2$} -- 5s5d\,{$^3$D$_{1,2,3}$} spectroscopy}

\begin{table*}
		\begin{tabular*}{\textwidth}{@{\extracolsep{\fill}}lclccccc}    \hline\hline \noalign{\smallskip}
&   &   & 5s5p\,{$^3$P$_2$} -- 5s5d\,{$^3$D$_1$}   &  5s5p\,{$^3$P$_2$} -- 5s5d\,{$^3$D$_2$}   &  5s5p\,{$^3$P$_2$} -- 5s5d\,{$^3$D$_3$} \\
\noalign{\smallskip}\hline\noalign{\smallskip}
$^{84}$Sr & & & 602\,833\,830.9 (2.0)       & 603\,285\,766.0 (2.0)     & 603\,976\,415.0 (2.0)  \\
$^{86}$Sr & & & 602\,833\,876.1 (2.0)       & 603\,285\,811.3 (2.0)     & 603\,976\,460.0 (2.0)  \\
$^{88}$Sr & & & 602\,833\,924.9 (2.0)       & 603\,285\,858.8 (2.0)     & 603\,976\,506.6 (2.0)  \\
\noalign{\smallskip}\hline\noalign{\smallskip}
$^{87}$Sr & $5/2 \rightarrow 3/2$ & &        					& 				   	& 603\,976\,697.5 (2.0)  \\
$^{87}$Sr & $5/2 \rightarrow 5/2$ & &       					& 603\,284\,256.9 (2.0)   		& -  \\
$^{87}$Sr & $5/2 \rightarrow 7/2$ & & 602\,830\,264.7 (2.0)     		& 603\,284\,005.0 (2.0)   		& -  \\
$^{87}$Sr & $7/2 \rightarrow 5/2$ & &                         			& 603\,285\,034.2 (2.0)          	& -  \\
$^{87}$Sr & $7/2 \rightarrow 7/2$ & & 602\,831\,041.6 (2.0)       	& 603\,284\,772.0 (2.0)   		& -  \\
$^{87}$Sr & $7/2 \rightarrow 9/2$ & &   -    					& 603\,284\,459.8 (2.0)   		& -  \\
$^{87}$Sr & $9/2 \rightarrow 7/2$ & & 602\,832\,017.1 (2.0)      		& 603\,285\,748.6 (2.0)      	& -  \\
$^{87}$Sr & $9/2 \rightarrow 9/2$ & & 602\,833\,045.1 (2.0)       	& 603\,285\,436.5 (2.0)   		& -  \\
$^{87}$Sr & $9/2 \rightarrow 11/2$ & & 602\,834\,291.5 (2.0)      	& 603\,285\,047.3 (2.0)   		& -  \\
$^{87}$Sr & $11/2 \rightarrow 9/2$ & & 602\,834\,207.3 (2.0)      	&  603\,286\,608.8 (2.0)  		& -  \\
$^{87}$Sr & $11/2 \rightarrow 11/2$ & & 602\,835\,459.1 (2.0)   	&  603\,286\,211.1 (2.0)  		& -  \\
$^{87}$Sr & $11/2 \rightarrow 13/2$ & &                           			&  603\,285\,749.4 (2.0)     	& -  \\
$^{87}$Sr & $13/2 \rightarrow 11/2$ & & 602\,836\,808.7 (2.0)    	& 603\,287\,563.6 (2.0)   		& 603\,978\,453 (10)  \\
$^{87}$Sr & $13/2 \rightarrow 13/2$ & &       					& 603\,287\,099.1 (2.0)     	& 603\,977\,428 (10)  \\
$^{87}$Sr & $13/2 \rightarrow 15/2$ & &      					& 					& 603\,976\,258.8 (2.0) \\ \hline \hline
		\end{tabular*}
    \caption{Transition frequencies, given in MHz, of the three 5s5p\,{$^3$P$_2$} -- 5s5d\,{$^3$D$_{1,2,3}$} ``green'' transitions around 497\,nm. Dashes indicate transitions that are allowed by selection rules, but were not observed.}
    \label{tab:green_data}
\end{table*}

We will now present spectroscopic data of the``green'' 5s5p\,{$^3$P$_2$ -- 5s5d\,{$^3$D$_{1,2,3}$} transition around 497\,nm, using absorption imaging as the detection method. The loading time of the reservoir is adjusted to compensate for the different natural abundances of the bosonic isotopes, and the intensity of the repump light is $5\times 10^{-4}\, I_{\mathrm{sat}}$. A typical set of measurements is shown in Fig.~\ref{fig:fig4}. Panel (a) shows the spectra of the three bosonic isotopes, where the isotope shifts are roughly 20\,MHz per mass unit.

Panel (b) shows measurements on the fermionic isotope, taken with a typical intensity of $1\times 10^{-3}\, I_{\mathrm{sat}}$. Note that the spectrum spans many GHz in frequency, and that the amplitude of the signal rises more than two orders of magnitude above the background. Note also the two overlapping lines around 603\,285\,750\,MHz, which were discussed in the previous section and in Fig.~\ref{fig:fig3}(e).

A compilation of the transition frequencies can be found in Tab.~\ref{tab:green_data}. As discussed in the Appendix, the absolute frequencies have an error of 2\,MHz. Relative frequency differences, such as the isotope shifts, have an uncertainty of 200\,kHz. The measurement of the hyperfine structure of the fermionic isotope is more involved, leading to a larger error in the magnetic dipole and electric quadrupole interaction constants $A$ and $Q$.  Compared to spectroscopy in a MOT or discharge, the spectroscopy scheme used here can improve both the precision and accuracy by at least two orders of magnitude.

\subsection{5s5p\,{$^3$P$_2$} -- 5s6d\,{$^3$D$_{1,2,3}$} spectroscopy}

We will now introduce a potential technological simplification to strontium experiments that employ the strategy of accumulation and subsequent repumping. Instead of using the $^3$S$_1$ state, which requires two lasers for repumping, such experiments might choose a state that requires only one repump laser, \textit{e.g.}~one of the 5s$n$d\,{$^3$D$_2$} states. Experiments so far used the state with $n=4$, which requires a technically involved laser source at $3.01\,\mu$m, or the state with $n=5$ at 497\,nm, which requires a frequency-doubled diode laser system. As a potential simplification, we explore the use of blue light at 403\,nm, employing the state with $n=6$ for repumping; see Fig.~\ref{fig:fig1}. Diode lasers at this wavelength are readily available thanks to Blu-ray technology.

As a first step to characterize this approach, we perform reservoir spectroscopy on this transition, analogous to the measurements described above. These measurements are plaqued by frequency noise of the laser, which is not locked to a stable reference. This frequency noise occurs on the timescale of a few 100\,ms, with an amplitude of a few MHz. We expect this noise to be the dominant error source and therefore match our detection scheme to the less stringent requirements in the precision. Specifically, we do not apply the experimentally involved capture into the narrow-line MOT and subsequent absorption imaging, but recapture into the broad-line MOT and use the peak in MOT fluorescence as a measure of the number of repumped atoms; see Sec.~\ref{sec:fluorescence}. A compilation of the resonance positions is given in Tab.~\ref{tab:blue_data}.

\begin{table*}
		\begin{tabular*}{\textwidth}{@{\extracolsep{\fill}}lclccccc}    \hline\hline \noalign{\smallskip}
&   &   &  5s5p\,{$^3$P$_2$} -- 5s5d\,{$^3$D$_1$}   &  5s5p\,{$^3$P$_2$} -- 5s5d\,{$^3$D$_2$}  &  5s5p\,{$^3$P$_2$} -- 5s5d\,{$^3$D$_3$} \\
\noalign{\smallskip}\hline\noalign{\smallskip}
$^{84}$Sr & & & 743\,105\,100 (5)       & 743\,252\,879 (2)     & 743\,621\,723 (5)  \\
$^{86}$Sr & & & 743\,105\,147 (5)       & 743\,252\,911 (2)     & 743\,621\,767 (5)  \\
$^{88}$Sr & & & 743\,105\,177 (5)       & 743\,252\,960 (2)     & 743\,621\,808 (5)  \\
\noalign{\smallskip}\hline\noalign{\smallskip}
$^{87}$Sr & $5/2 \rightarrow 3/2$ & &        				& 				& 743\,622\,062 (5)  \\
$^{87}$Sr & $5/2 \rightarrow 5/2$ & &       				& -   	                    		& 743\,621\,642 (5)  \\
$^{87}$Sr & $5/2 \rightarrow 7/2$ & & 743\,101\,462 (5)      	& -                     		& -  \\
$^{87}$Sr & $7/2 \rightarrow 5/2$ & &                         		& -          			& -  \\
$^{87}$Sr & $7/2 \rightarrow 7/2$ & & 743\,102\,252 (5)       	& 743\,252\,566 (5)     	& -  \\
$^{87}$Sr & $7/2 \rightarrow 9/2$ & &   -    				& -   	                   		& -  \\
$^{87}$Sr & $9/2 \rightarrow 7/2$ & & 743\,103\,222 (5)      	& 743\,253\,506 (5)         	& -  \\
$^{87}$Sr & $9/2 \rightarrow 9/2$ & & 743\,104\,293 (5)       	& 743\,252\,802 (5)     	& -  \\
$^{87}$Sr & $9/2 \rightarrow 11/2$ & & 743\,105\,617 (5)    	& 743\,251\,903 (5)     	& -  \\
$^{87}$Sr & $11/2 \rightarrow 9/2$ & & 743\,105\,456 (5)    	&  743\,253\,983 (5)    	& 743\,623\,285 (5) \\
$^{87}$Sr & $11/2 \rightarrow 11/2$ & & 743\,106\,788 (5)    	&  743\,253\,090 (5)   	& 743\,622\,387 (5) \\
$^{87}$Sr & $11/2 \rightarrow 13/2$ & &                           		&  743\,252\,000 (5)      	& 743\,621\,332 (5) \\
$^{87}$Sr & $13/2 \rightarrow 11/2$ & & 743\,108\,137 (5)     	& 743\,254\,431 (5)        	& 743\,623\,722 (5) \\
$^{87}$Sr & $13/2 \rightarrow 13/2$ & &       				& 743\,253\,352 (5)         	& 743\,622\,677 (5) \\
$^{87}$Sr & $13/2 \rightarrow 15/2$ & &      				& 				& 743\,621\,475 (5) \\ \hline \hline
		\end{tabular*}
    \caption{Transition frequencies, given in MHz, of the three 5s5p\,{$^3$P$_2$} -- 5s6d\,{$^3$D$_{1,2,3}$} ``blue'' transitions around 403\,nm. Dashes indicate transitions that are allowed by selection rules, but were not observed.}
    \label{tab:blue_data}
\end{table*}

The relevant parameters of the three 5s5p\,{$^3$P$_2$} -- 5s$n$d\,{$^3$D$_J$}, $n \in (4,5,6)$, repump transitions are listed in Tab.~\ref{tab:Summary}.

\subsection{Hyperfine structure}

In contrast to the bosonic isotopes with zero nuclear spin, the fermionic strontium isotope has $I=9/2$ and shows an involved hyperfine structure. For the $^3$P$_2$ -- $^3$D$_J$ transitions of  $^{87}$Sr, we can identify each of the resonances by their relative positions, and subsequently calculate the interaction constants $A$ and $Q$ of the $^3$D$_{1,2}$ states using
\begin{equation*}
    \Delta E_{\mathrm{hfs}}/h = \frac{A}{2} K + \frac{Q}{2}\frac{\frac{3}{4} K(K+1)-I(I+1)J(J+1)}{I(2I-1)J(2J-1)}.
\end{equation*}
Here, $K=F(F+1)-I(I+1)-J(J+1)$, where $I=9/2$ is the nuclear spin, $J$ is the total angular momentum, and $F$ denotes the hyperfine state, given by $\vec{F}=\vec{I}+\vec{J}$. The hyperfine structure of the $^3$P$_2$ state is known to the kHz-level \cite{Heider1977hso} ($A=-212.765(1)$ and $Q=67.215(15)$), and we fix these values for our calculation. The hyperfine structure of the 5s4d\,{$^3$D$_J$} states has been determined independently \cite{Bushaw1993hsi}. A qualitative comparison with our values assures that our designation of hyperfine states is correct.

The $^3$P$_2$ -- $^3$D$_3$} transitions have the lowest repumping efficiency, as the probability of atoms in the $^3$D$_3$ state to decay into the $^3$P$_1$ state is relatively low; see Secs.~\ref{sec:efficiency} and \ref{sec:cycling}. Instead, atoms might decay into another $^3$P$_2$ hyperfine state, where they are dark for the repump light. For the green 5s5p\,{$^3$P$_2$} -- 5s5d\,{$^3$D$_3$} transition, only four out of 15 lines could be detected, despite a careful search over more than 8\,GHz. We speculate that the two strongest lines are the cycling $F=5/2 \rightarrow F'=3/2$ and $F=13/2 \rightarrow F'=15/2$ transitions, which would imply that the hyperfine structure of the $^3$D$_3$ state is inverted with an interaction constant $A\approx -150$. Simultaneous repumping from various hyperfine states would be necessary to obtain a complete spectrum of the $^3$P$_2$ -- $^3$D$_3$ hyperfine transitions. 

The signal strength on the blue 5s5p\,{$^3$P$_2$} -- 5s6d\,{$^3$D$_3$} transition is much larger compared to the green transition, allowing us to detect significantly more lines. The reason is an increased branching ratio into the $^3$P$_1$ state; see Sec.~\ref{sec:efficiency}. The eight resonances have very different amplitudes, which helps us to identify them through our knowledge the atoms' initial distribution over the $^3$P$_2$ hyperfine states. 

\begin{table*}
		\begin{tabular*}{\textwidth}{@{\extracolsep{\fill}}llccccccccc}    \hline\hline \noalign{\smallskip}
\multicolumn{1}{c}{transition} &     & $\lambda$ [nm]   & $\nu(^{88}$Sr) [MHz]   & $\Delta_{84}^{88}$ [MHz]  & $\Delta_{86}^{88}$ [MHz] & $\Delta_{87}^{88}$ [MHz] & $A$ [MHz]  & $Q$ [MHz]\\
\noalign{\smallskip}\hline\noalign{\smallskip}
5s5p\,{$^3$P$_2$} -- 5s4d\,{$^3$D$_1$} & & 3067.0  \cite{Sansonetti2010wtp}   & \hspace{1.5mm}97\,747\,180 (150) \cite{Sansonetti2010wtp}    & 	   &        &     & \hspace{2.5mm}139.9 (2)  \cite{Bushaw1993hsi}   & 15 (2)  \cite{Bushaw1993hsi}\\
5s5p\,{$^3$P$_2$} -- 5s4d\,{$^3$D$_2$} & & 3011.8 \cite{Mickelson2009ras}    & \hspace{1.5mm}99\,537\,870 (75) \, \cite{Mickelson2009ras}   & 600 (50) \cite{Mickelson2009ras}  & 270 (40) \cite{Mickelson2009ras} & 110 (30) \cite{Mickelson2009ras} & $-78.08$ (5)  \cite{Bushaw1993hsi}  & 18 (1)   \cite{Bushaw1993hsi}\\
5s5p\,{$^3$P$_2$} -- 5s4d\,{$^3$D$_3$} & & 2923.4 \cite{Sansonetti2010wtp}  & 102\,550\,490 (170) \cite{Sansonetti2010wtp}   &          &          &        & $-115.3$ (2)    \cite{Bushaw1993hsi}   &  51 (9)  \cite{Bushaw1993hsi}\\
\noalign{\smallskip}\hline\noalign{\smallskip}
5s5p\,{$^3$P$_2$} -- 5s5d\,{$^3$D$_1$} & & 497.30     & 602\,833\,924.9 (2.0)   & 94.0 (2)   & 48.8 (2)  & 38 (2)  & \hspace{1mm}227.3 (7)     	& 0 (10)  \\
5s5p\,{$^3$P$_2$} -- 5s5d\,{$^3$D$_2$} & & 496.93     & 603\,285\,858.8 (2.0)   & 92.8 (2)   & 47.5 (2)  & 17 (2)  & $-71.5$ (5)   	& 0 (30)  \\
5s5p\,{$^3$P$_2$} -- 5s5d\,{$^3$D$_3$} & & 496.36     & 603\,976\,506.6 (2.0)   & 91.6 (2)   & 46.6 (2)  & \hspace{1.5mm}27 (2)$^{\rm a}$  & $-156.9$ (3)$^{\rm a}$	&  \hspace{1.5mm}0 (30)$^{\rm a}$ \\
\noalign{\smallskip}\hline\noalign{\smallskip}
5s5p\,{$^3$P$_2$} -- 5s6d\,{$^3$D$_1$} & & 403.43     & 743\,105\,177 (5)    & 77 (10)   & 30 (10)   & 18 (2)  	& \hspace{2.5mm}239.7 (5)     	& \hspace{1.5mm}5 (20)   \\
5s5p\,{$^3$P$_2$} -- 5s6d\,{$^3$D$_2$} & & 403.35     & 743\,252\,960 (2)    & 81 (2)\hspace{1.5mm}     & 49 (2)\hspace{1.5mm}     & 20 (5)  	& $-163.2$ (9)  	& 30 (20) \\
5s5p\,{$^3$P$_2$} -- 5s6d\,{$^3$D$_3$} & & 403.15     & 743\,621\,808 (5)    & 85 (10)   & 41 (10)   & 53 (5) 	& $-161.8$ (5)      	& 20 (20) \\ \hline \hline
		\end{tabular*}
\caption{Compilation of relevant spectroscopic data. For each transition, we state the wavelength $\lambda$ and frequency $\nu$ of the most abundant $^{88}$Sr isotope, as well as the isotope shifts $\Delta_{84}^{88} = \nu(^{88}{\rm Sr}) - \nu(^{84}{\rm Sr})$, and $\Delta_{86}^{88} = \nu(^{88}{\rm Sr}) - \nu(^{86}{\rm Sr})$, and $\Delta_{87}^{88} = \nu(^{88}{\rm Sr}) - \nu(^{87}{\rm Sr})$. The interaction constants $A$ and $Q$ characterize the hyperfine structure of the $^3$D$_J$ manifolds in the fermionic isotope $^{87}$Sr. Values labelled by $^{\rm a}$ are based on a speculative assignment of the transitions; see Sec.~\ref{sec:data} C.}
\label{tab:Summary}
\end{table*}

\section{Repump efficiency}
\label{sec:efficiency}

\begin{figure}
\includegraphics[width=\columnwidth]{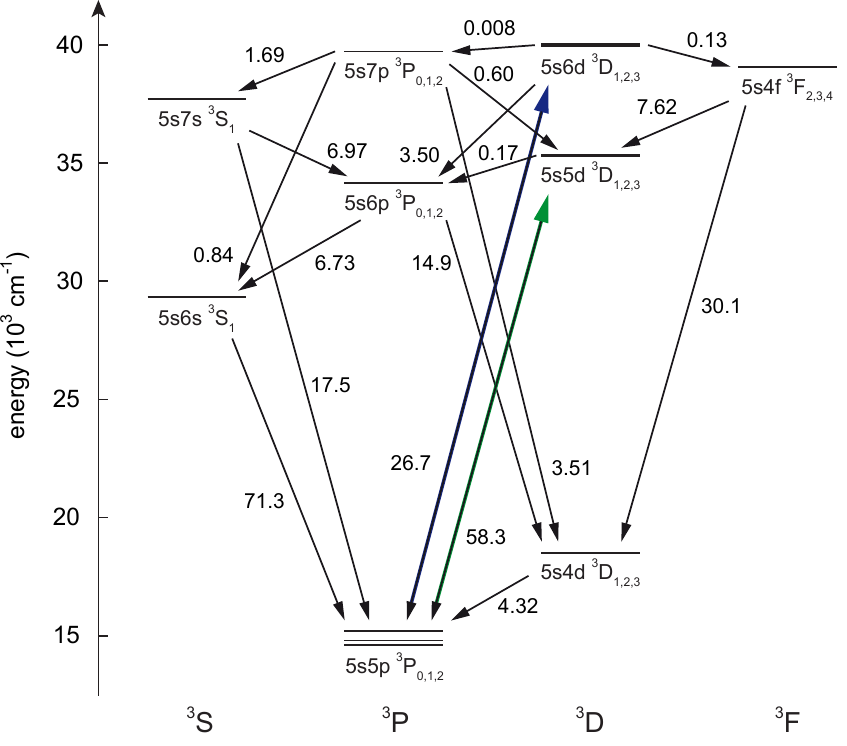}
\caption{Transition probabilities $A_{ik}$, given in units of $10^6\,{\rm s}^{-1}$, between triplet states of Sr relevant for this work. Values are taken from Ref.~\cite{Werij1992osa}. The fine structure splitting, indicated here only for the lowest $^3$P$_J$ states, is otherwise smaller than the thickness of the horizontal lines. Decay paths with negligible contribution are omitted.}
\label{fig:fig5}
\end{figure}

The efficiency of optical pumping can be reduced by decay from the excited state into undesired states. For the specific case studied here, atoms in the $^3$P$_2$ state are to be pumped into the $^3$P$_1$ state. The undesired state is the $^3$P$_0$ state, which is very long-lived, not trapped in the magnetic trap, and dark to repump photons through the finestructure splitting of about $600\,{\rm cm}^{-1}$. Clearly, the probability of decay into the $^3$P$_0$ state depends on the excited state used for repumping, and might assume any value between zero and close to unity.

The 5s6s\,{$^3$S$_1$} state can decay into the $^3$P$_0$ state: the branching ratios into the $^3$P$_{0,1,2}$ states are $1/9$, $3/9$, and $5/9$, respectively, such that 25\% of all repumped atoms end up in the $^3$P$_0$ state. To close this leak, strontium experiments operate another repumper at 679\,nm to excite these atoms again into the $^3$S$_1$ state \cite{Dinneen1999cco}.

The 5s4d\,{$^3$D$_2$} state can only decay into the 5s5p\,{$^3$P$_{1,2}$} states due to selection rules. Decay into the $^3$P$_0$ state is possible only through higher-order processes, which will be neglected here. When going up the ladder of $^3$D$_2$ states, however, more and more additional states appear into which the $^3$D$_2$ state can decay; see Fig.~\ref{fig:fig5}. These might have nonzero branching ratios into the $^3$P$_0$ state, possibly through a cascade of various other states.

We will now quantify the branching ratio of higher-lying 5s$n$d\,{$^3$D$_{1,2,3}$} states into the three 5s5p\,{$^3$P$_{0,1,2}$} states. We perform a detailed calculation to trace all possible decay channels for the cases of $n \in (5,6)$, as illustrated in Fig.~\ref{fig:fig5}. We use the transition probabilities from Ref.~\cite{Werij1992osa}. Again, we will consider only dipole-allowed (E1) transitions. The branching ratios into the different fine structure states are obtained using the standard Wigner 6-$j$ symbols. We will consider only bosonic isotopes of zero nuclear spin here, ignoring the hyperfine structure of the fermionic isotope. Similarly, we have no knowledge about the $m_J$-state distribution of atoms in the 5s$n$d\,{$^3$D$_{1,2,3}$} states, and therefore neglect a possible $m_J$-state dependence of the branching ratios. The fine structure splitting of the states involved is a few $100\,{\rm cm}^{-1}$ for the 5s5p\,{$^3$P$_{0,1,2}$} states and much smaller for all other states. We take these splittings into account by multiplying correction factors to the branching ratios \cite{Ludlow2008tso}. The correction factors favor decay into fine structure states of lower energy (lower $J$ quantum number) by typically a few percent. We carefully compare our results with the transition probabilities tabulated in Ref.~\cite{Sansonetti2010wtp} and find an overall agreement within a few percent.

\begin{table*}
\begin{tabular*}{\textwidth}{@{\extracolsep{\fill}}cccccccccc}    \hline\hline \noalign{\smallskip}
state used     & \hspace{3mm}  & \multicolumn{3}{c}{calculated branching ratios}   & \hspace{7mm} & \multicolumn{4}{c}{repump efficiency}  \\
for repumping  & & 5s5p\,{$^3$P$_0$}   &  5s5p\,{$^3$P$_1$}   &  5s5p\,{$^3$P$_2$} & & calculated   & via $\tau_{\rm rMOT}$ & via $N_{\rm rMOT}$ & via $N_{\rm repump}$ \\ \hline \noalign{\smallskip}

5s5d\,{$^3$D$_1$} &   &  56.24\%       & 41.11\%            & 2.644\% &    & 42.22\%     &  		&           		& 	             	 \\
5s5d\,{$^3$D$_2$} &   &  0.035\%       & 76.00\%            & 23.96\%  &   & 99.95\%     & 96.9(5)\%	& 97(1)\%		& 100\%      	 \\
5s5d\,{$^3$D$_3$} &   &  0.012\%       & 0.057\%            & 99.93\%   &  & 82.61\%     & 		&           		& 	             	  \\ \hline \noalign{\smallskip}
5s6d\,{$^3$D$_1$} &   &  53.20\%       & 41.75\%            & 5.071\%  &   & 43.98\%     & 40(5)\%	& 40(2)\%		&           		  \\
5s6d\,{$^3$D$_2$} &   &  1.363\%       & 72.80\%            & 25.83\%  &   & 98.16\%     & 95.7(9)\%	& 92(1)\%		& 98.1(5)\%  	  \\
5s6d\,{$^3$D$_3$} &   &  0.467\%       & 2.276\%            & 97.25\%   &  & 82.97\%     & 84(2)\%	& 79(1)\%		& 82.6(5)\%  	  \\ \hline \hline
\end{tabular*}
\caption{Branching ratios of relevant $^3$D$_2$ states into the 5s5p\,{$^3$P$_{0,1,2}$} states, taking into account all possible decay paths. The precision is given by the uncertainty of the transition probabilities taken from Ref.~\cite{Werij1992osa}; see the text for details. The right part of the table shows the repump efficiency of each of the six transitions, as calculated from the branching ratios and experimentally determined from the lifetime $\tau_{\rm rMOT}$, the number of atoms in the repumped MOT $N_{\rm rMOT}$, and the number of atoms repumped form the reservoir $N_{\rm repump}$. The values of the last column were normalized to their largest value.}
\label{tab:tab_branching}
\end{table*}

The results of these calculations are presented in Tab.~\ref{tab:tab_branching}. We find that repumping on the green 5s5p\,{$^3$P$_2$} -- 5s5d\,{$^3$D$_2$} transition induces a decay into the 5s5p\,{$^3$P$_0$} state of about 0.05\%. The multitude of higher-lying states increases this value to about 1.8\% for the blue 5s5p\,{$^3$P$_2$} -- 5s6d\,{$^3$D$_2$} transition. These values, albeit small, are significant for continuous repumping in the MOT.

We will now follow three different experimental approaches to quantify the relative branching ratio from the 5s5d\,{$^3$D$_2$} state into the 5s5p\,{$^3$P$_{0,1}$} states. These approaches boil down to a precise measurement of $\tau_{\rm rMOT}$, $N_{\rm rMOT}$, and $N_{\rm repump}$; see Fig.~\ref{fig:fig2}. At first, we measure the lifetime of the bosonic $^{88}$Sr MOT, with and without the repumper applied. The MOT lifetime is strongly density-dependent, therefore measurements are taken with a rather dilute MOT far from saturation. We measure a lifetime of 30(5)\,ms without repumper. This lifetime increases to $\tau_{\rm rMOT} = 950(50)\,$ms with the green repumper applied, and to 700(50)\,ms with the blue repumper. The repumpers close the leak into the $^3$P$_2$ state, and we assume that residual decay into the $^3$P$_0$ state limits the lifetime of the repumped MOTs. The vacuum lifetime, measured with atoms held in an optical dipole trap, is about 2 minutes. The factor of $32(5)$ in lifetime increase for the green transition (factor 23(5) for the blue transition) directly shows that about 3\% of the repump cycles using green light, and about 4\% of the repump cycles using blue light lead the atom into the dark $^3$P$_0$ state. Corresponding values of further transitions are given in the column labelled ``via $\tau_{\rm rMOT}$'' in Tab.~\ref{tab:tab_branching}.

In a second approach, we measure the steady-state fluorescence of a weakly charged MOT at low MOT light intensity, corresponding to $N_{\rm rMOT}$. When applying the green repumper, we measure a fluorescence increase by a factor of 30(2) \cite{Sorrentino2006lca}, and a factor of 12(1) for the blue repumper. These numbers can be translated into a probability of atoms to decay from the $^3$D$_2$ state into the $^3$P$_0$ state. This probability is about 3\% for the green repumper and about 8\% for the blue repumper; see the column labelled ``via $N_{\rm rMOT}$'' in Tab.~\ref{tab:tab_branching}.

All of these experimentally determined values are significantly larger than the ones calculated above. One possible explanation is a different decay channel not considered so far:~the pathways $^1$P$_1$ $\rightarrow$ $^3$P$_0$ and $^1$P$_1$ $\rightarrow$ $^3$D$_1$ $\rightarrow$ $^3$P$_0$. The estimated branching ratio from the $^1$P$_1$ state into the 5s4d\,{$^3$D$_{J}$} states is $2.8 \times 10^{-7}$ \cite{Bidel2002per}, about two orders of magnitude smaller than the decay into the 5s4d\,{$^1$D$_2$} state \cite{Werij1992osa}. We estimate that about one third of the atoms following this pathway end up in the dark $^3$P$_0$ state. The probability of the direct loss channel  $^1$P$_1$ $\rightarrow$ $^3$P$_0$ is not known. Given the uncertainties in the estimated transition probability values, we find that this decay mechanism might very well explain our observed values of  $\tau_{\rm rMOT}$ and $N_{\rm rMOT}$. This hypothesis is strengthened by the fact that experiments employing repumping of both the $^3$P$_0$ and $^3$P$_2$ states have reported MOT lifetimes of up to 10 seconds \cite{MartinPrivComm}. The branching ratio from the $^1$P$_1$ state via all pathways into the $^3$P$_0$ state would then be $1:1.7(3) \times 10^6$.

We perform a third experiment that does not include the direct decay of the $^1$P$_1$ state into the metastable states. Here, we directly measure the number $N_{\rm repump}$ of atoms repumped from the reservoir. We assume that repumping on the green 5s5p\,{$^3$P$_2$} -- 5s5d\,{$^3$D$_2$} transition returns all atoms into the ground state, and we use this value to normalize the performance of two other transitions; see last column of Tab.~\ref{tab:tab_branching}. We find the number of atoms repumped on the blue 5s5p\,{$^3$P$_2$\ -- 5s6p\,{$^3$D$_2$} transition to be reduced by 1.9\% compared to repumping on the green 5s5p\,{$^3$P$_2$} -- 5s5p\,{$^3$D$_2$} transition, in very good agreement with our calculation.

Two conclusions can be drawn from these measurements: (1) Using the 5s6d\,{$^3$D$_2$} state instead of the 5s5d\,{$^3$D$_2$} is 1.9\% less efficient for a single repump cycle, which seems tolerable when repumping previously accumulated atoms from the reservoir. This is very promising, as it offers a substantial simplification of the required laser system. (2) The direct decay from the $^1$P$_1$ state into the metastable states seems to be significantly larger than expected, and sophisticated calculations might help to resolve this issue. A countinuously repumped MOT will therefore greatly benefit in atom number from repumping of the $^3$P$_0$ state. Usage of the $^3$S$_1$ state here might require the least technological effort.

\section{Is the 5s5p\,{$^3$P$_2$} -- 5s5d\,{$^3$D$_3$} transition cycling?}
\label{sec:cycling}

\begin{figure}
\includegraphics[width=\columnwidth]{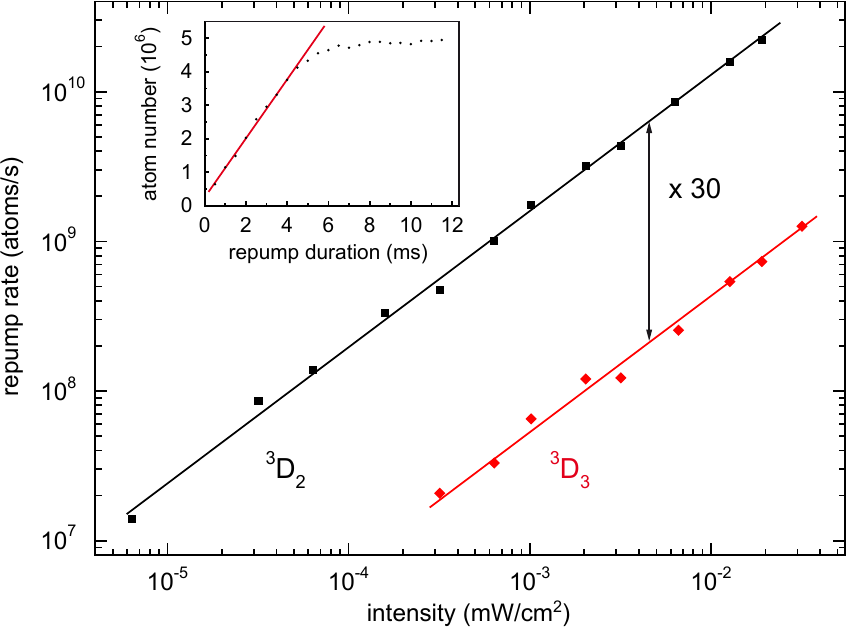}
\caption{Repump rate in dependence of light intensity. The number of atoms repumped into the $^1$S$_0$ ground state is proportional to the repump time before reaching saturation (inset, taken on the green 5s5p\,{$^3$P$_2$} -- 5s5d\,{$^3$D$_3$} transition at $I=1\times10^{-2}\,I_{\mathrm{sat}}$). This rate is measured for a large range of intensities on both of the green $^3$P$_2$ -- $^3$D$_2$ and $^3$P$_2$ -- $^3$D$_3$ transitions.}
\label{fig:fig6}
\end{figure}

The 5s5p\,{$^3$P$_2$} -- 5s$n$d{$^3$D$_3$} transitions are unique in that they are the only (nearly) cycling transitions originating from the $^3$P$_J$ manifold. There is an infinite ladder of 5s$n$d\,{$^3$D$_3$} states reachable from the $^3$P$_2$ state, starting from $n=4$. 

These transitions have been proposed as MOT transitions to cool atoms in the metastable $^3$P$_2$ state. Doppler cooling on similar transitions has been performed in calcium \cite{Grunert2002sdm} and neon \cite{Feldker2011mot}. The 5s5p\,{$^3$P$_2$} -- 5s4d\,{$^3$D$_3$} transition at $2.92\,\mu$m is predestined for operation of a MOT due to a combination of a large wavelength and a small linewidth of only 50\,kHz, leading to both a low Doppler and a low recoil temperature. In contrast to all higher-lying $^3$D$_3$ states, no other states appear in between the two MOT states to spoil cycling operation, and the absorption of two photons does not ionize the atom. Apart from the fact that lasers at this frequency and linewidth are difficult to operate, efficient MOT operation might however be hindered by inelastic losses \cite{Traverso2009iae}.

The $^3$P$_2$ -- $^3$D$_3$ transitions could also be used for fluorescence and absorption imaging. A recent publication on quantum computation with fermionic $^{87}$Sr \cite{Daley2008qcw} suggests the use of a $^3$P$_2$ -- $^3$D$_3$, $|F=13/2\rangle \rightarrow |F'=15/2\rangle$ transition for fluorescence readout of atoms in the metastable states. This application would benefit from a short-wavelength transition, which brings about good imaging resolution and large detection efficiency. The 5s5p\,{$^3$P$_2$} -- 5s5d\,{$^3$D$_3$} transition at 497\,nm seems a promising candidate. Analogous to the discussion in Sec.~\ref{sec:efficiency}, this transition might not be perfectly cycling due to decay paths that lead into the $^3$P$_{0,1}$ states. These additional channels gain importance when climbing up the ladder of 5s$n$d\,{$^3$D$_3$} states.

 In the following, we will investigate the probability of atoms in the $^3$D$_3$ state to decay into the 5s5p\,{$^3$P$_{0,1}$} states. A summation of all possible decay paths for two $^3$D$_3$ states can be found in Tab.~\ref{tab:tab_branching}. The branching ratio into the $^3$P$_1$ state is 0.057\% for the 5s5d\,{$^3$D$_3$} state and 2.28\% for the 5s6d\,{$^3$D$_3$} state. Again, we neglected higher-order transitions. These numbers already suggest that fluorescence detection of individual atoms will require optical pumping of the $^3$P$_{0,1}$ states into the $^3$P$_2$ state, as on the order of $10^5$ photons need to be cycled for a sufficiently large signal. For the fermionic isotope, the optical pumping scheme will be more involved due to the hyperfine structure and multitude of $m_F$ states.

To verify our calculation, we perform an experiment that allows us to compare the decay rates from the 5s5d\,{$^3$D$_2$} and $^3$D$_3$ states into the 5s5p\,{$^3$P$_1$} state. The 5s5p\,{$^3$P$_2$} -- 5s5d\,{$^3$D$_3$} transition is certainly not completely cycling, as can be inferred from the spectroscopy described in Sec.~\ref{sec:data}: obviously, some atoms are transferred into the $^3$P$_1$ state and further into the $^1$S$_0$ ground state; see Fig.~\ref{fig:fig5}. The measurements presented in the following are performed with the bosonic $^{88}$Sr isotope. About $10^7$ atoms are loaded into the $^3$P$_2$ reservoir state, repumped, and recaptured in the narrow-line MOT. As a reference, we first measure the number of atoms repumped on the 5s5p\,{$^3$P$_2$} -- 5s5d\,{$^3$D$_2$} transition in dependence of repump time. For low intensities and short repumping times, there is a linear dependence of repumped atoms on repump time until the atom number saturates; see the inset of Fig.~\ref{fig:fig6}. We perform this measurement for a wide range of repump light intensities and, for each measurement series, extract the initial repump rate.

We then perform the same experiment on the $^3$P$_2$ -- $^3$D$_3$ transition; see Fig.~\ref{fig:fig6}. We find that repumping via the $^3$D$_3$ state instead of the $^3$D$_2$ state returns 30(1)-times less atoms into the ground state. The saturation intensities are 2.2\,mW/cm$^2$ for the $^3$P$_2$ -- $^3$D$_2$ and 10.4\,mW/cm$^2$ for the $^3$P$_2$ -- $^3$D$_3$ transition. From the calculated branching ratios to the $^3$P$_1$ state (see Tab.~\ref{tab:tab_branching}), we expect a ratio of  $76/0.057 \approx 1300$ between the repumping rates of the two transitions examined, constituting a clear discrepancy.

The enhanced decay of the $^3$D$_3$ state into the $^3$P$_1$ state could be explained by the presence of the black-body radiation (BBR) field. Many of the transitions between the higher-lying states have wavelengths comparable to the maximum of the BBR at room temperature, which is around $10\,\mu$m. Such transitions can be driven by the BBR field. As a coincidence, the energy difference of 1049\,cm$^{-1}$ between the 5s6p\,{$^3$P$_2$} and 5s5d\,{$^3$D$_3$} states corresponds exactly to the maximum of the room-temperature BBR spectrum. It has been observed in other systems that BBR can redistribute population between near-degenerate levels. Such a process was crucial for the operation of a Ra MOT \cite{Guest2007lto}, and we speculate about a similar process for our case. Similar processes might also contribute to the unexpectedly low repump efficiencies discussed in Sec.~\ref{sec:efficiency}.

\section{Conclusion}
\label{sec:conclusion}

In conclusion, we have performed spectroscopy on 5s5p\,{$^3$P$_2$} -- 5s$n$d\,{$^3$D$_{1,2,3}$} transitions in strontium and determined the isotope shifts and hyperfine parameters. We have accessed all relevant $^3$D$_J$ $\rightarrow$ $^3$P$_J$ branching ratios both theoretically and experimentally. The results have immediate implications for the usage of these transitions in protocols of cooling, manipulation, and detection of ultracold strontium atoms.

The general scheme of performing spectroscopy on atoms stored in a metastable reservoir state can be applied to many other species as well. Elements such as magnesium, calcium \cite{Hansen2005dfs}, ytterbium, and mercury have an electronic structure very similar to strontium. More complex atoms, such as the rare-earth species \cite{Hancox2004mto} dysprosium \cite{Lu2010tud,Lu2011sdb}, holmium \cite{Miao2014mot}, erbium \cite{McClelland2006lcw,Aikawa2012bec}, and thulium \cite{Sukachev2010mot}, which are currently under investigation, all have a wealth of metastable states which can be trapped in the magnetic field of a MOT and potentially be used for reservoir spectroscopy.

\begin{acknowledgments}

We thank R.~Grimm for generous support and stimulating discussions. We thank M.~Chwalla, C.~Roos and R.~Blatt for the loan of the wavemeter and for providing us with an absolute frequency reference based on the $^{40}\text{Ca}^+$ ion. We highly appreciate extensive technical assistance by C.~Hempel. We also wish to thank the strontium groups at JILA, LENS, and Rice University for sharing their experience with us. We gratefully acknowledge support from the Austrian Ministry of Science and Research (BMWF) and the Austrian Science Fund (FWF) through a START grant under project number Y507-N20. As member of the project iSense, we also acknowledge the financial support of the Future and Emerging Technologies (FET) programme within the Seventh Framework Programme for Research of the European Commission, under FET-Open grant No.~250072.

\end{acknowledgments}

\section*{Appendix: Error analysis}
\label{sec:appendix}

\textit{Laser at 497\,nm} --- The spectroscopy light at 497\,nm is generated by a frequency-doubled diode laser, which delivers about 40\,mW of green light. The infrared laser is locked to a stable reference cavity, which has a drift of typically 100\,kHz per hour, mainly due to piezo creep and possibly thermal drift. The green light is passed through a cascade of four acousto-optic modulators (AOMs) in double-pass configuration. The AOMs have a design frequency of 350\,MHz and a bandwidth of about 150\,MHz, resulting in an overall scan range of 1.2\,GHz. The RF used to drive the AOMs is generated by direct digital synthesizers (DDSs), which are referenced to better than 1\,Hz to either the global positioning system GPS or a commercial rubidium atomic clock, both of which have negligible error. The laser light itself has a linewidth smaller than 1\,MHz. About 1\,mW of light is available for the experiment and collimated to a beam of 10\,mm diameter to interrogate the atoms.

\textit{Laser at 403\,nm} --- The spectroscopy light at 403\,nm is generated by a diode laser, emitting about 30\,mW. The light is sent through a series of two 350-MHz AOMs in double-pass configuration and delivered via a fiber to the experiment, where the beam is expanded to illuminate the entire MOT volume. This laser is not locked to a high-finesse cavity, and it shows frequency noise on timescales of a few 100\,ms, maximum excursions being about 2\,MHz. The drift rate of the laser is rather poor, typically 100\,MHz per hour.

\textit{Absolute frequency} --- The frequency of the respective laser used for spectroscopy is constantly monitored on a commercial High Finesse WSU/2 wavemeter. The wavemeter is constantly calibrated by a Ti:Saph laser at 729\,nm. This laser is locked to a high-finesse cavity with a well-characterized drift rate of about 200\,Hz per hour. The absolute frequency is determined by spectroscopy of the 4s\,$^2$S$_{1/2}$ -- 3d\,$^2$D$_{5/2}$ transition in the $^{40}\mathrm{Ca}^+$ ion at 432\,042\,129\,776\,393.2\,(1.0)\,Hz \cite{Chwalla2009afm}. Calibration of the wavemeter has an accuracy of 500\,kHz, which can be reduced substantially by averaging over many calibrations. If not constantly re-calibrated, the wavemeter drifts by about 100\,kHz per hour. The precision of the wavemeter is specified to be about 1\,MHz.

The accuracy of the wavemeter is about 1\,MHz if used within a few nm of the calibration wavelength, but might be substantially larger further away from the calibration wavelength. To characterize the calibration fidelity of the wavemeter over a large range of wavelengths, we measure the frequency of various lasers referenced to various well-established optical transitions, of which the absolute frequencies are well known from the literature. More specifically, these are the 4s\,{$^2$S$_{1/2}$} -- 4p\,{$^2$P$_{1/2}$} dipole transition at 397\,nm in the $^{40}$Ca$^+$ ion, which is known to within 1.7\,MHz \cite{Wolf2008fmo}; the 5s$^2$\,{$^1$S$_0$} -- 5s5p\,{$^3$P$_1$} intercombination line at 689\,nm in atomic $^{88}$Sr, which is known to within 10\,kHz \cite{Ferrari2003pfm,Courtillot2005aso}; and the 5s\,{$^2$S$_{1/2}$} -- 5p\,{$^2$P$_{3/2}$} dipole transition at 780\,nm in atomic $^{87}$Rb, which is known to within 6\,kHz \cite{Ye1996hsa}. The infrared light at 793\,nm, which is used to generate the light at 397\,nm, is used as well. The deviations of the wavemeter reading from the real frequency scatter within about $\pm5\,$MHz; see Fig.~\ref{fig:fig7}. We use our frequency-doubled lasers at 397/793\,nm, 461/922\,nm, and 497/994\,nm to connect between different wavelength ranges, as the frequency conversion between the fundamental and the doubled frequency is known to be exactly 2 by design. We perform additional consistency checks with light at wavelengths 461\,nm, 854\,nm, and 866\,nm, corresponding to dipole transitions in either $^{40}$Ca$^+$ or $^{88}$Sr, which are known at a level of order 100\,MHz. From this characterization, we conclude that the wavemeter is accurate to within at least 10\,MHz over the entire range from 400 to 1000\,nm. As a spin-off, we can determine the frequency of the 461\,nm transition in $^{88}$Sr to be 650\,503\,775 (10)\,MHz, corresponding to an energy of the 5s5p\,{$^1$P$_1$} state of 21\,698.4703(3)\,cm$^{-1}$. This value is more precise than (but in slight disagreement with) the one given in Ref.~\cite{Sansonetti2010wtp}.

\begin{figure}[t]
\includegraphics[width=\columnwidth]{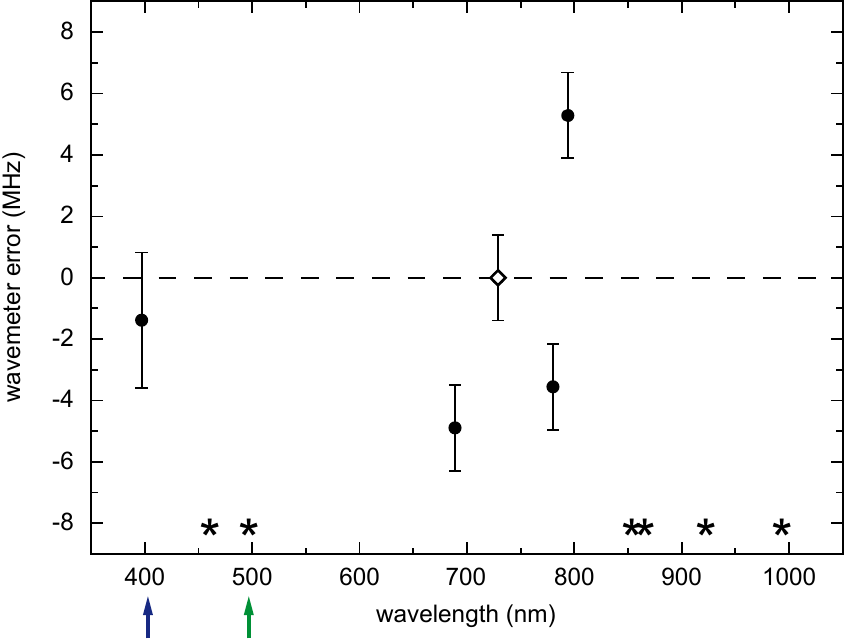}
\caption{Characterization of the wavemeter error. The wavemeter is calibrated at 729\,nm (open diamond) and used to measure light at various well-known frequencies (filled circles). We plot the difference between the obtained values and the literature values, where the error bars are dominated by the calibration and measurement uncertainty of the wavemeter.  The stars denote wavelengths at which additional consistency checks were performed, however at a smaller precision. The spectroscopy measurements presented in this work were performed at 403 and 497\,nm, denoted by arrows.}
\label{fig:fig7}
\end{figure}

\textit{Error in absolute frequencies} --- The errors given in Tabs.~\ref{tab:green_data} and \ref{tab:blue_data} include solely the precision of the wavemeter and the uncertainty in the fit to the data points, but neglect the inaccuracy of the wavemeter, which is certainly below 10\,MHz; see Fig.~\ref{fig:fig7}. The error of the absolute frequencies is then entirely dominated by the inaccuracy of the wavemeter. The data is fit by a Lorentzian profile, where deviations from a Lorentzian lineshape are visible only for extremely low repump intensities. The error in the determination of the centriod ranges between 50\,kHz and a few MHz, depending on the signal strength. This error becomes comparable to the wavemeter uncertainty only for very weak transitions.

\textit{Error in relative frequencies} --- The error in frequency differences, such as isotope shifts, can be substantially smaller, as the inaccuracy of the wavemeter drops out. For the measurement of the isotope shifts between the bosons, the wavemeter is carefully calibrated and then left free-running. A slow drift of the wavemeter then constitutes the dominant error source. The complete spectroscopy scan of the bosonic isotopes is performed within less than half an hour, in which the wavemeter drifts by much less than 100\,kHz. A complete scan of all the hyperfine transitions of the fermionic isotope takes about a day, and the wavemeter might drift by up to a few MHz in this period. As a conservative estimate, the isotope shifts between the bosonic isotopes can be measured with an uncertainty of about 200\,kHz, and the hyperfine splittings can be determined to about 4\,MHz. While the values stated here apply for the green transition, another error source appears for the blue transition:~The frequency noise of the laser, which is about 1\,MHz on timescales below 1\,s, leads to the comparably large errors given in Tab.~\ref{tab:Summary}.

\textit{Stability and reproducibility} --- We carefully determine the stability of our MOT atom number and find it to be better than 1\% on all timescales between a millisecond and many hours. During a measurement campaign, we repetitively measure certain transitions over the course of a few days to exclude any drifts and find the scatter in absolute frequency to be within less than 1\,MHz. Selected frequencies were determined in two independent measurement campaigns separated by a time interval of two years and using largely different setups, and the frequency difference was found to be within the 1\,MHz uncertainty of the wavemeter, thereby excluding a large number of possible systematic errors.

\bibliographystyle{apsrev}

%\bibliography{spectroscopy,ultracold}

\begin{thebibliography}{100}
\expandafter\ifx\csname natexlab\endcsname\relax\def\natexlab#1{#1}\fi
\expandafter\ifx\csname bibnamefont\endcsname\relax
  \def\bibnamefont#1{#1}\fi
\expandafter\ifx\csname bibfnamefont\endcsname\relax
  \def\bibfnamefont#1{#1}\fi
\expandafter\ifx\csname citenamefont\endcsname\relax
  \def\citenamefont#1{#1}\fi
\expandafter\ifx\csname url\endcsname\relax
  \def\url#1{\texttt{#1}}\fi
\expandafter\ifx\csname urlprefix\endcsname\relax\def\urlprefix{URL }\fi
\providecommand{\bibinfo}[2]{#2}
\providecommand{\eprint}[2][]{\url{#2}}

\bibitem[{\citenamefont{Ido and Katori}(2003)}]{Ido2003rfs}
\bibinfo{author}{\bibfnamefont{T.}~\bibnamefont{Ido}} \bibnamefont{and}
  \bibinfo{author}{\bibfnamefont{H.}~\bibnamefont{Katori}},
  \bibinfo{journal}{Phys. Rev. Lett.} \textbf{\bibinfo{volume}{91}},
  \bibinfo{pages}{053001} (\bibinfo{year}{2003}).

\bibitem[{\citenamefont{Takamoto et~al.}(2005)\citenamefont{Takamoto, Hong,
  Higashi, and Katori}}]{Takamoto2005aol}
\bibinfo{author}{\bibfnamefont{M.}~\bibnamefont{Takamoto}},
  \bibinfo{author}{\bibfnamefont{F.-L.} \bibnamefont{Hong}},
  \bibinfo{author}{\bibfnamefont{R.}~\bibnamefont{Higashi}}, \bibnamefont{and}
  \bibinfo{author}{\bibfnamefont{H.}~\bibnamefont{Katori}},
  \bibinfo{journal}{Nature} \textbf{\bibinfo{volume}{435}},
  \bibinfo{pages}{321} (\bibinfo{year}{2005}).

\bibitem[{\citenamefont{Derevianko and Katori}(2011)}]{Derevianko2011poo}
\bibinfo{author}{\bibfnamefont{A.}~\bibnamefont{Derevianko}} \bibnamefont{and}
  \bibinfo{author}{\bibfnamefont{H.}~\bibnamefont{Katori}},
  \bibinfo{journal}{Rev. Mod. Phys.} \textbf{\bibinfo{volume}{83}},
  \bibinfo{pages}{331} (\bibinfo{year}{2011}).

\bibitem[{\citenamefont{Hinkley et~al.}(2013)\citenamefont{Hinkley, Sherman,
  Phillips, Schioppo, Lemke, Beloy, Pizzocaro, Oates, and
  Ludlow}}]{Hinkley2013aac}
\bibinfo{author}{\bibfnamefont{N.}~\bibnamefont{Hinkley}},
  \bibinfo{author}{\bibfnamefont{J.~A.} \bibnamefont{Sherman}},
  \bibinfo{author}{\bibfnamefont{N.~B.} \bibnamefont{Phillips}},
  \bibinfo{author}{\bibfnamefont{M.}~\bibnamefont{Schioppo}},
  \bibinfo{author}{\bibfnamefont{N.~D.} \bibnamefont{Lemke}},
  \bibinfo{author}{\bibfnamefont{K.}~\bibnamefont{Beloy}},
  \bibinfo{author}{\bibfnamefont{M.}~\bibnamefont{Pizzocaro}},
  \bibinfo{author}{\bibfnamefont{C.~W.} \bibnamefont{Oates}}, \bibnamefont{and}
  \bibinfo{author}{\bibfnamefont{A.~D.} \bibnamefont{Ludlow}},
  \bibinfo{journal}{Science} \textbf{\bibinfo{volume}{341}},
  \bibinfo{pages}{1215} (\bibinfo{year}{2013}).

\bibitem[{\citenamefont{Bloom et~al.}(2014)\citenamefont{Bloom, Nicholson,
  Williams, Campbell, Bishof, Zhang, Zhang, Bromley, and Ye}}]{Bloom2014aol}
\bibinfo{author}{\bibfnamefont{B.~J.} \bibnamefont{Bloom}},
  \bibinfo{author}{\bibfnamefont{T.~L.} \bibnamefont{Nicholson}},
  \bibinfo{author}{\bibfnamefont{J.~R.} \bibnamefont{Williams}},
  \bibinfo{author}{\bibfnamefont{S.~L.} \bibnamefont{Campbell}},
  \bibinfo{author}{\bibfnamefont{M.}~\bibnamefont{Bishof}},
  \bibinfo{author}{\bibfnamefont{X.}~\bibnamefont{Zhang}},
  \bibinfo{author}{\bibfnamefont{W.}~\bibnamefont{Zhang}},
  \bibinfo{author}{\bibfnamefont{S.~L.} \bibnamefont{Bromley}},
  \bibnamefont{and} \bibinfo{author}{\bibfnamefont{J.}~\bibnamefont{Ye}},
  \bibinfo{journal}{Nature} \textbf{\bibinfo{volume}{506}}, \bibinfo{pages}{71}
  (\bibinfo{year}{2014}).

\bibitem[{\citenamefont{Poli et~al.}(2011)\citenamefont{Poli, Wang, Tarallo,
  Alberti, Prevedelli, and Tino}}]{Poli2011pmo}
\bibinfo{author}{\bibfnamefont{N.}~\bibnamefont{Poli}},
  \bibinfo{author}{\bibfnamefont{F.-Y.} \bibnamefont{Wang}},
  \bibinfo{author}{\bibfnamefont{M.~G.} \bibnamefont{Tarallo}},
  \bibinfo{author}{\bibfnamefont{A.}~\bibnamefont{Alberti}},
  \bibinfo{author}{\bibfnamefont{M.}~\bibnamefont{Prevedelli}},
  \bibnamefont{and} \bibinfo{author}{\bibfnamefont{G.~M.} \bibnamefont{Tino}},
  \bibinfo{journal}{Phys. Rev. Lett.} \textbf{\bibinfo{volume}{106}},
  \bibinfo{pages}{038501} (\bibinfo{year}{2011}).

\bibitem[{\citenamefont{Takasu et~al.}(2003)\citenamefont{Takasu, Maki, Komori,
  Takano, Honda, Kumakura, Yabuzaki, and Takahashi}}]{Takasu2003ssb}
\bibinfo{author}{\bibfnamefont{Y.}~\bibnamefont{Takasu}},
  \bibinfo{author}{\bibfnamefont{K.}~\bibnamefont{Maki}},
  \bibinfo{author}{\bibfnamefont{K.}~\bibnamefont{Komori}},
  \bibinfo{author}{\bibfnamefont{T.}~\bibnamefont{Takano}},
  \bibinfo{author}{\bibfnamefont{K.}~\bibnamefont{Honda}},
  \bibinfo{author}{\bibfnamefont{M.}~\bibnamefont{Kumakura}},
  \bibinfo{author}{\bibfnamefont{T.}~\bibnamefont{Yabuzaki}}, \bibnamefont{and}
  \bibinfo{author}{\bibfnamefont{Y.}~\bibnamefont{Takahashi}},
  \bibinfo{journal}{Phys. Rev. Lett.} \textbf{\bibinfo{volume}{91}},
  \bibinfo{pages}{040404} (\bibinfo{year}{2003}).

\bibitem[{\citenamefont{Fukuhara
  et~al.}(2007{\natexlab{a}})\citenamefont{Fukuhara, Sugawa, and
  Takahashi}}]{Fukuhara2007bec}
\bibinfo{author}{\bibfnamefont{T.}~\bibnamefont{Fukuhara}},
  \bibinfo{author}{\bibfnamefont{S.}~\bibnamefont{Sugawa}}, \bibnamefont{and}
  \bibinfo{author}{\bibfnamefont{Y.}~\bibnamefont{Takahashi}},
  \bibinfo{journal}{Phys. Rev. A} \textbf{\bibinfo{volume}{76}},
  \bibinfo{pages}{051604(R)} (\bibinfo{year}{2007}{\natexlab{a}}).

\bibitem[{\citenamefont{Fukuhara
  et~al.}(2007{\natexlab{b}})\citenamefont{Fukuhara, Takasu, Kumakura, and
  Takahashi}}]{Fukuhara2007dfg}
\bibinfo{author}{\bibfnamefont{T.}~\bibnamefont{Fukuhara}},
  \bibinfo{author}{\bibfnamefont{Y.}~\bibnamefont{Takasu}},
  \bibinfo{author}{\bibfnamefont{M.}~\bibnamefont{Kumakura}}, \bibnamefont{and}
  \bibinfo{author}{\bibfnamefont{Y.}~\bibnamefont{Takahashi}},
  \bibinfo{journal}{Phys. Rev. Lett.} \textbf{\bibinfo{volume}{98}},
  \bibinfo{pages}{030401} (\bibinfo{year}{2007}{\natexlab{b}}).

\bibitem[{\citenamefont{Fukuhara et~al.}(2009)\citenamefont{Fukuhara, Sugawa,
  Takasu, and Takahashi}}]{Fukuhara2009aof}
\bibinfo{author}{\bibfnamefont{T.}~\bibnamefont{Fukuhara}},
  \bibinfo{author}{\bibfnamefont{S.}~\bibnamefont{Sugawa}},
  \bibinfo{author}{\bibfnamefont{Y.}~\bibnamefont{Takasu}}, \bibnamefont{and}
  \bibinfo{author}{\bibfnamefont{Y.}~\bibnamefont{Takahashi}},
  \bibinfo{journal}{Phys. Rev. A} \textbf{\bibinfo{volume}{79}},
  \bibinfo{pages}{021601(R)} (\bibinfo{year}{2009}).

\bibitem[{\citenamefont{Sugawa et~al.}(2011)\citenamefont{Sugawa, Yamazaki,
  Taie, and Takahashi}}]{Sugawa2011bec}
\bibinfo{author}{\bibfnamefont{S.}~\bibnamefont{Sugawa}},
  \bibinfo{author}{\bibfnamefont{R.}~\bibnamefont{Yamazaki}},
  \bibinfo{author}{\bibfnamefont{S.}~\bibnamefont{Taie}}, \bibnamefont{and}
  \bibinfo{author}{\bibfnamefont{Y.}~\bibnamefont{Takahashi}},
  \bibinfo{journal}{Phys. Rev. A} \textbf{\bibinfo{volume}{84}},
  \bibinfo{pages}{011610} (\bibinfo{year}{2011}).

\bibitem[{\citenamefont{Kraft et~al.}(2009)\citenamefont{Kraft, Vogt, Appel,
  Riehle, and Sterr}}]{Kraft2009bec}
\bibinfo{author}{\bibfnamefont{S.}~\bibnamefont{Kraft}},
  \bibinfo{author}{\bibfnamefont{F.}~\bibnamefont{Vogt}},
  \bibinfo{author}{\bibfnamefont{O.}~\bibnamefont{Appel}},
  \bibinfo{author}{\bibfnamefont{F.}~\bibnamefont{Riehle}}, \bibnamefont{and}
  \bibinfo{author}{\bibfnamefont{U.}~\bibnamefont{Sterr}},
  \bibinfo{journal}{Phys. Rev. Lett.} \textbf{\bibinfo{volume}{103}},
  \bibinfo{pages}{130401} (\bibinfo{year}{2009}).

\bibitem[{\citenamefont{Halder et~al.}(2012)\citenamefont{Halder, Yang, and
  Hemmerich}}]{Halder2012art}
\bibinfo{author}{\bibfnamefont{P.}~\bibnamefont{Halder}},
  \bibinfo{author}{\bibfnamefont{C.-Y.} \bibnamefont{Yang}}, \bibnamefont{and}
  \bibinfo{author}{\bibfnamefont{A.}~\bibnamefont{Hemmerich}},
  \bibinfo{journal}{Phys. Rev. A} \textbf{\bibinfo{volume}{85}},
  \bibinfo{pages}{031603} (\bibinfo{year}{2012}).

\bibitem[{\citenamefont{Stellmer et~al.}(2009)\citenamefont{Stellmer, Tey,
  Huang, Grimm, and Schreck}}]{Stellmer2009bec}
\bibinfo{author}{\bibfnamefont{S.}~\bibnamefont{Stellmer}},
  \bibinfo{author}{\bibfnamefont{M.~K.} \bibnamefont{Tey}},
  \bibinfo{author}{\bibfnamefont{B.}~\bibnamefont{Huang}},
  \bibinfo{author}{\bibfnamefont{R.}~\bibnamefont{Grimm}}, \bibnamefont{and}
  \bibinfo{author}{\bibfnamefont{F.}~\bibnamefont{Schreck}},
  \bibinfo{journal}{Phys. Rev. Lett.} \textbf{\bibinfo{volume}{103}},
  \bibinfo{pages}{200401} (\bibinfo{year}{2009}).

\bibitem[{\citenamefont{{Martinez de Escobar}
  et~al.}(2009)\citenamefont{{Martinez de Escobar}, Mickelson, Yan, DeSalvo,
  Nagel, and Killian}}]{MartinezdeEscobar2009bec}
\bibinfo{author}{\bibfnamefont{Y.~N.} \bibnamefont{{Martinez de Escobar}}},
  \bibinfo{author}{\bibfnamefont{P.~G.} \bibnamefont{Mickelson}},
  \bibinfo{author}{\bibfnamefont{M.}~\bibnamefont{Yan}},
  \bibinfo{author}{\bibfnamefont{B.~J.} \bibnamefont{DeSalvo}},
  \bibinfo{author}{\bibfnamefont{S.~B.} \bibnamefont{Nagel}}, \bibnamefont{and}
  \bibinfo{author}{\bibfnamefont{T.~C.} \bibnamefont{Killian}},
  \bibinfo{journal}{Phys. Rev. Lett.} \textbf{\bibinfo{volume}{103}},
  \bibinfo{pages}{200402} (\bibinfo{year}{2009}).

\bibitem[{\citenamefont{Mickelson et~al.}(2010)\citenamefont{Mickelson,
  Martinez~de Escobar, Yan, DeSalvo, and Killian}}]{Mickelson2010bec}
\bibinfo{author}{\bibfnamefont{P.~G.} \bibnamefont{Mickelson}},
  \bibinfo{author}{\bibfnamefont{Y.~N.} \bibnamefont{Martinez~de Escobar}},
  \bibinfo{author}{\bibfnamefont{M.}~\bibnamefont{Yan}},
  \bibinfo{author}{\bibfnamefont{B.~J.} \bibnamefont{DeSalvo}},
  \bibnamefont{and} \bibinfo{author}{\bibfnamefont{T.~C.}
  \bibnamefont{Killian}}, \bibinfo{journal}{Phys. Rev. A}
  \textbf{\bibinfo{volume}{81}}, \bibinfo{pages}{051601}
  (\bibinfo{year}{2010}).

\bibitem[{\citenamefont{DeSalvo et~al.}(2010)\citenamefont{DeSalvo, Yan,
  Mickelson, {Martinez de Escobar}, and Killian}}]{DeSalvo2010dfg}
\bibinfo{author}{\bibfnamefont{B.~J.} \bibnamefont{DeSalvo}},
  \bibinfo{author}{\bibfnamefont{M.}~\bibnamefont{Yan}},
  \bibinfo{author}{\bibfnamefont{P.~G.} \bibnamefont{Mickelson}},
  \bibinfo{author}{\bibfnamefont{Y.~N.} \bibnamefont{{Martinez de Escobar}}},
  \bibnamefont{and} \bibinfo{author}{\bibfnamefont{T.~C.}
  \bibnamefont{Killian}}, \bibinfo{journal}{Phys. Rev. Lett.}
  \textbf{\bibinfo{volume}{105}}, \bibinfo{pages}{030402}
  (\bibinfo{year}{2010}).

\bibitem[{\citenamefont{Tey et~al.}(2010)\citenamefont{Tey, Stellmer, Grimm,
  and Schreck}}]{Tey2010ddb}
\bibinfo{author}{\bibfnamefont{M.~K.} \bibnamefont{Tey}},
  \bibinfo{author}{\bibfnamefont{S.}~\bibnamefont{Stellmer}},
  \bibinfo{author}{\bibfnamefont{R.}~\bibnamefont{Grimm}}, \bibnamefont{and}
  \bibinfo{author}{\bibfnamefont{F.}~\bibnamefont{Schreck}},
  \bibinfo{journal}{Phys. Rev. A} \textbf{\bibinfo{volume}{82}},
  \bibinfo{pages}{011608} (\bibinfo{year}{2010}).

\bibitem[{\citenamefont{Stellmer et~al.}(2010)\citenamefont{Stellmer, Tey,
  Grimm, and Schreck}}]{Stellmer2010bec}
\bibinfo{author}{\bibfnamefont{S.}~\bibnamefont{Stellmer}},
  \bibinfo{author}{\bibfnamefont{M.~K.} \bibnamefont{Tey}},
  \bibinfo{author}{\bibfnamefont{R.}~\bibnamefont{Grimm}}, \bibnamefont{and}
  \bibinfo{author}{\bibfnamefont{F.}~\bibnamefont{Schreck}},
  \bibinfo{journal}{Phys. Rev. A} \textbf{\bibinfo{volume}{82}},
  \bibinfo{pages}{041602} (\bibinfo{year}{2010}).

\bibitem[{\citenamefont{Stellmer et~al.}(2013)\citenamefont{Stellmer, Grimm,
  and Schreck}}]{Stellmer2013poq}
\bibinfo{author}{\bibfnamefont{S.}~\bibnamefont{Stellmer}},
  \bibinfo{author}{\bibfnamefont{R.}~\bibnamefont{Grimm}}, \bibnamefont{and}
  \bibinfo{author}{\bibfnamefont{F.}~\bibnamefont{Schreck}},
  \bibinfo{journal}{Phys. Rev. A} \textbf{\bibinfo{volume}{87}},
  \bibinfo{pages}{013611} (\bibinfo{year}{2013}).

\bibitem[{\citenamefont{Cazalilla and Rey}(2014)}]{Cazalilla2014ufg}
\bibinfo{author}{\bibfnamefont{M.~A.} \bibnamefont{Cazalilla}}
  \bibnamefont{and} \bibinfo{author}{\bibfnamefont{A.~M.} \bibnamefont{Rey}},
  \bibinfo{journal}{arXiv:1402.2792}  (\bibinfo{year}{2014}).

\bibitem[{\citenamefont{Wu et~al.}(2003)\citenamefont{Wu, Hu, and
  Zhang}}]{Wu2003ess}
\bibinfo{author}{\bibfnamefont{C.}~\bibnamefont{Wu}},
  \bibinfo{author}{\bibfnamefont{J.-P.} \bibnamefont{Hu}}, \bibnamefont{and}
  \bibinfo{author}{\bibfnamefont{S.-C.} \bibnamefont{Zhang}},
  \bibinfo{journal}{Phys. Rev. Lett.} \textbf{\bibinfo{volume}{91}},
  \bibinfo{pages}{186402} (\bibinfo{year}{2003}).

\bibitem[{\citenamefont{Wu}(2006)}]{Wu2006hsa}
\bibinfo{author}{\bibfnamefont{C.}~\bibnamefont{Wu}}, \bibinfo{journal}{Mod.
  Phys. Lett. B} \textbf{\bibinfo{volume}{20}}, \bibinfo{pages}{1707}
  (\bibinfo{year}{2006}).

\bibitem[{\citenamefont{Cazalilla et~al.}(2009)\citenamefont{Cazalilla, Ho, and
  Ueda}}]{Cazalilla2009ugo}
\bibinfo{author}{\bibfnamefont{M.~A.} \bibnamefont{Cazalilla}},
  \bibinfo{author}{\bibfnamefont{A.}~\bibnamefont{Ho}}, \bibnamefont{and}
  \bibinfo{author}{\bibfnamefont{M.}~\bibnamefont{Ueda}}, \bibinfo{journal}{New
  J. Phys.} \textbf{\bibinfo{volume}{11}}, \bibinfo{pages}{103033}
  (\bibinfo{year}{2009}).

\bibitem[{\citenamefont{Hermele et~al.}(2009)\citenamefont{Hermele, Gurarie,
  and Rey}}]{Hermele2009mio}
\bibinfo{author}{\bibfnamefont{M.}~\bibnamefont{Hermele}},
  \bibinfo{author}{\bibfnamefont{V.}~\bibnamefont{Gurarie}}, \bibnamefont{and}
  \bibinfo{author}{\bibfnamefont{A.~M.} \bibnamefont{Rey}},
  \bibinfo{journal}{Phys. Rev. Lett.} \textbf{\bibinfo{volume}{103}},
  \bibinfo{pages}{135301} (\bibinfo{year}{2009}).

\bibitem[{\citenamefont{Gorshkov et~al.}(2010)\citenamefont{Gorshkov, Hermele,
  Gurarie, Xu, Julienne, Ye, Zoller, Demler, Lukin, and Rey}}]{Gorshkov2010tos}
\bibinfo{author}{\bibfnamefont{A.}~\bibnamefont{Gorshkov}},
  \bibinfo{author}{\bibfnamefont{M.}~\bibnamefont{Hermele}},
  \bibinfo{author}{\bibfnamefont{V.}~\bibnamefont{Gurarie}},
  \bibinfo{author}{\bibfnamefont{C.}~\bibnamefont{Xu}},
  \bibinfo{author}{\bibfnamefont{P.}~\bibnamefont{Julienne}},
  \bibinfo{author}{\bibfnamefont{J.}~\bibnamefont{Ye}},
  \bibinfo{author}{\bibfnamefont{P.}~\bibnamefont{Zoller}},
  \bibinfo{author}{\bibfnamefont{E.}~\bibnamefont{Demler}},
  \bibinfo{author}{\bibfnamefont{M.~D.} \bibnamefont{Lukin}}, \bibnamefont{and}
  \bibinfo{author}{\bibfnamefont{A.~M.} \bibnamefont{Rey}},
  \bibinfo{journal}{Nature Phys.} \textbf{\bibinfo{volume}{6}},
  \bibinfo{pages}{289} (\bibinfo{year}{2010}).

\bibitem[{\citenamefont{Foss-Feig et~al.}(2010)\citenamefont{Foss-Feig,
  Hermele, and Rey}}]{FossFeig2010ptk}
\bibinfo{author}{\bibfnamefont{M.}~\bibnamefont{Foss-Feig}},
  \bibinfo{author}{\bibfnamefont{M.}~\bibnamefont{Hermele}}, \bibnamefont{and}
  \bibinfo{author}{\bibfnamefont{A.~M.} \bibnamefont{Rey}},
  \bibinfo{journal}{Phys. Rev. A} \textbf{\bibinfo{volume}{81}},
  \bibinfo{pages}{051603} (\bibinfo{year}{2010}).

\bibitem[{\citenamefont{Xu}(2010)}]{Xu2010lim}
\bibinfo{author}{\bibfnamefont{C.}~\bibnamefont{Xu}}, \bibinfo{journal}{Phys.
  Rev. B} \textbf{\bibinfo{volume}{81}}, \bibinfo{pages}{144431}
  (\bibinfo{year}{2010}).

\bibitem[{\citenamefont{Hung et~al.}(2011)\citenamefont{Hung, Wang, and
  Wu}}]{Hung2011qmi}
\bibinfo{author}{\bibfnamefont{H.-H.} \bibnamefont{Hung}},
  \bibinfo{author}{\bibfnamefont{Y.}~\bibnamefont{Wang}}, \bibnamefont{and}
  \bibinfo{author}{\bibfnamefont{C.}~\bibnamefont{Wu}}, \bibinfo{journal}{Phys.
  Rev. B} \textbf{\bibinfo{volume}{84}}, \bibinfo{pages}{054406}
  (\bibinfo{year}{2011}).

\bibitem[{\citenamefont{Taie et~al.}(2010)\citenamefont{Taie, Takasu, Sugawa,
  Yamazaki, Tsujimoto, Murakami, and Takahashi}}]{Taie2010roa}
\bibinfo{author}{\bibfnamefont{S.}~\bibnamefont{Taie}},
  \bibinfo{author}{\bibfnamefont{Y.}~\bibnamefont{Takasu}},
  \bibinfo{author}{\bibfnamefont{S.}~\bibnamefont{Sugawa}},
  \bibinfo{author}{\bibfnamefont{R.}~\bibnamefont{Yamazaki}},
  \bibinfo{author}{\bibfnamefont{T.}~\bibnamefont{Tsujimoto}},
  \bibinfo{author}{\bibfnamefont{R.}~\bibnamefont{Murakami}}, \bibnamefont{and}
  \bibinfo{author}{\bibfnamefont{Y.}~\bibnamefont{Takahashi}},
  \bibinfo{journal}{Phys. Rev. Lett.} \textbf{\bibinfo{volume}{105}},
  \bibinfo{pages}{190401} (\bibinfo{year}{2010}).

\bibitem[{\citenamefont{Taie et~al.}(2012)\citenamefont{Taie, Yamazaki, Sugawa,
  and Takahashi}}]{Taie2012asm}
\bibinfo{author}{\bibfnamefont{S.}~\bibnamefont{Taie}},
  \bibinfo{author}{\bibfnamefont{R.}~\bibnamefont{Yamazaki}},
  \bibinfo{author}{\bibfnamefont{S.}~\bibnamefont{Sugawa}}, \bibnamefont{and}
  \bibinfo{author}{\bibfnamefont{Y.}~\bibnamefont{Takahashi}},
  \bibinfo{journal}{Nat Phys} \textbf{\bibinfo{volume}{8}},
  \bibinfo{pages}{825} (\bibinfo{year}{2012}).

\bibitem[{\citenamefont{Martin et~al.}(2013)\citenamefont{Martin, Bishof,
  Swallows, Zhang, Benko, von Stecher, Gorshkov, Rey, and Ye}}]{Martin2013aqm}
\bibinfo{author}{\bibfnamefont{M.~J.} \bibnamefont{Martin}},
  \bibinfo{author}{\bibfnamefont{M.}~\bibnamefont{Bishof}},
  \bibinfo{author}{\bibfnamefont{M.~D.} \bibnamefont{Swallows}},
  \bibinfo{author}{\bibfnamefont{X.}~\bibnamefont{Zhang}},
  \bibinfo{author}{\bibfnamefont{C.}~\bibnamefont{Benko}},
  \bibinfo{author}{\bibfnamefont{J.}~\bibnamefont{von Stecher}},
  \bibinfo{author}{\bibfnamefont{A.~V.} \bibnamefont{Gorshkov}},
  \bibinfo{author}{\bibfnamefont{A.~M.} \bibnamefont{Rey}}, \bibnamefont{and}
  \bibinfo{author}{\bibfnamefont{J.}~\bibnamefont{Ye}},
  \bibinfo{journal}{Science} \textbf{\bibinfo{volume}{341}},
  \bibinfo{pages}{632} (\bibinfo{year}{2013}).

\bibitem[{\citenamefont{Zhang et~al.}(2014)\citenamefont{Zhang, Bishof,
  Bromley, Kraus, Safronova, Zoller, Rey, and Ye}}]{Zhang2014soo}
\bibinfo{author}{\bibfnamefont{X.}~\bibnamefont{Zhang}},
  \bibinfo{author}{\bibfnamefont{M.}~\bibnamefont{Bishof}},
  \bibinfo{author}{\bibfnamefont{S.~L.} \bibnamefont{Bromley}},
  \bibinfo{author}{\bibfnamefont{C.~V.} \bibnamefont{Kraus}},
  \bibinfo{author}{\bibfnamefont{M.~S.} \bibnamefont{Safronova}},
  \bibinfo{author}{\bibfnamefont{P.}~\bibnamefont{Zoller}},
  \bibinfo{author}{\bibfnamefont{A.~M.} \bibnamefont{Rey}}, \bibnamefont{and}
  \bibinfo{author}{\bibfnamefont{J.}~\bibnamefont{Ye}},
  \bibinfo{journal}{arXiv:1403.2964}  (\bibinfo{year}{2014}).

\bibitem[{\citenamefont{Scazza et~al.}(2014)\citenamefont{Scazza, Hofrichter,
  H\"ofer, De~Groot, Bloch, and F\"olling}}]{Scazza2014oot}
\bibinfo{author}{\bibfnamefont{F.}~\bibnamefont{Scazza}},
  \bibinfo{author}{\bibfnamefont{C.}~\bibnamefont{Hofrichter}},
  \bibinfo{author}{\bibfnamefont{M.}~\bibnamefont{H\"ofer}},
  \bibinfo{author}{\bibfnamefont{P.~C.} \bibnamefont{De~Groot}},
  \bibinfo{author}{\bibfnamefont{I.}~\bibnamefont{Bloch}}, \bibnamefont{and}
  \bibinfo{author}{\bibfnamefont{S.}~\bibnamefont{F\"olling}},
  \bibinfo{journal}{arXiv:1403.4761}  (\bibinfo{year}{2014}).

\bibitem[{\citenamefont{Gerbier and Dalibard}(2010)}]{Gerbier2009gff}
\bibinfo{author}{\bibfnamefont{F.}~\bibnamefont{Gerbier}} \bibnamefont{and}
  \bibinfo{author}{\bibfnamefont{J.}~\bibnamefont{Dalibard}},
  \bibinfo{journal}{New J. Phys.} \textbf{\bibinfo{volume}{12}},
  \bibinfo{pages}{033007} (\bibinfo{year}{2010}).

\bibitem[{\citenamefont{Cooper}(2011)}]{Cooper2011ofl}
\bibinfo{author}{\bibfnamefont{N.~R.} \bibnamefont{Cooper}},
  \bibinfo{journal}{Phys. Rev. Lett.} \textbf{\bibinfo{volume}{106}},
  \bibinfo{pages}{175301} (\bibinfo{year}{2011}).

\bibitem[{\citenamefont{B\'eri and Cooper}(2011)}]{Beri2011zti}
\bibinfo{author}{\bibfnamefont{B.}~\bibnamefont{B\'eri}} \bibnamefont{and}
  \bibinfo{author}{\bibfnamefont{N.~R.} \bibnamefont{Cooper}},
  \bibinfo{journal}{Phys. Rev. Lett.} \textbf{\bibinfo{volume}{107}},
  \bibinfo{pages}{145301} (\bibinfo{year}{2011}).

\bibitem[{\citenamefont{G{\'o}recka et~al.}(2011)\citenamefont{G{\'o}recka,
  Gr\'emaud, and Miniatura}}]{Gorecka2011smf}
\bibinfo{author}{\bibfnamefont{A.}~\bibnamefont{G{\'o}recka}},
  \bibinfo{author}{\bibfnamefont{B.}~\bibnamefont{Gr\'emaud}},
  \bibnamefont{and}
  \bibinfo{author}{\bibfnamefont{C.}~\bibnamefont{Miniatura}},
  \bibinfo{journal}{Phys. Rev. A} \textbf{\bibinfo{volume}{84}},
  \bibinfo{pages}{023604} (\bibinfo{year}{2011}).

\bibitem[{\citenamefont{Dalibard et~al.}(2011)\citenamefont{Dalibard, Gerbier,
  Juzeliunas, and \"Ohberg}}]{Dalibard2011agp}
\bibinfo{author}{\bibfnamefont{J.}~\bibnamefont{Dalibard}},
  \bibinfo{author}{\bibfnamefont{F.}~\bibnamefont{Gerbier}},
  \bibinfo{author}{\bibfnamefont{G.}~\bibnamefont{Juzeliunas}},
  \bibnamefont{and} \bibinfo{author}{\bibfnamefont{P.}~\bibnamefont{\"Ohberg}},
  \bibinfo{journal}{Rev. Mod. Phys.} \textbf{\bibinfo{volume}{83}},
  \bibinfo{pages}{1523} (\bibinfo{year}{2011}).

\bibitem[{\citenamefont{Banerjee et~al.}(2013)\citenamefont{Banerjee, B\"ogli,
  Dalmonte, Rico, Stebler, Wiese, and Zoller}}]{Banerjee2013aqs}
\bibinfo{author}{\bibfnamefont{D.}~\bibnamefont{Banerjee}},
  \bibinfo{author}{\bibfnamefont{M.}~\bibnamefont{B\"ogli}},
  \bibinfo{author}{\bibfnamefont{M.}~\bibnamefont{Dalmonte}},
  \bibinfo{author}{\bibfnamefont{E.}~\bibnamefont{Rico}},
  \bibinfo{author}{\bibfnamefont{P.}~\bibnamefont{Stebler}},
  \bibinfo{author}{\bibfnamefont{U.-J.} \bibnamefont{Wiese}}, \bibnamefont{and}
  \bibinfo{author}{\bibfnamefont{P.}~\bibnamefont{Zoller}},
  \bibinfo{journal}{Phys. Rev. Lett.} \textbf{\bibinfo{volume}{110}},
  \bibinfo{pages}{125303} (\bibinfo{year}{2013}).

\bibitem[{\citenamefont{Diehl et~al.}(2010)\citenamefont{Diehl, Yi, Daley, and
  Zoller}}]{Diehl2010did}
\bibinfo{author}{\bibfnamefont{S.}~\bibnamefont{Diehl}},
  \bibinfo{author}{\bibfnamefont{W.}~\bibnamefont{Yi}},
  \bibinfo{author}{\bibfnamefont{A.~J.} \bibnamefont{Daley}}, \bibnamefont{and}
  \bibinfo{author}{\bibfnamefont{P.}~\bibnamefont{Zoller}},
  \bibinfo{journal}{Phys. Rev. Lett.} \textbf{\bibinfo{volume}{105}},
  \bibinfo{pages}{227001} (\bibinfo{year}{2010}).

\bibitem[{\citenamefont{Bhongale et~al.}(2013)\citenamefont{Bhongale, Mathey,
  Zhao, Yelin, and Lemeshko}}]{Bhongale2013qpo}
\bibinfo{author}{\bibfnamefont{S.~G.} \bibnamefont{Bhongale}},
  \bibinfo{author}{\bibfnamefont{L.}~\bibnamefont{Mathey}},
  \bibinfo{author}{\bibfnamefont{E.}~\bibnamefont{Zhao}},
  \bibinfo{author}{\bibfnamefont{S.~F.} \bibnamefont{Yelin}}, \bibnamefont{and}
  \bibinfo{author}{\bibfnamefont{M.}~\bibnamefont{Lemeshko}},
  \bibinfo{journal}{Phys. Rev. Lett.} \textbf{\bibinfo{volume}{110}},
  \bibinfo{pages}{155301} (\bibinfo{year}{2013}).

\bibitem[{\citenamefont{Olmos et~al.}(2013)\citenamefont{Olmos, Yu, Singh,
  Schreck, Bongs, and Lesanovsky}}]{Olmos2013lri}
\bibinfo{author}{\bibfnamefont{B.}~\bibnamefont{Olmos}},
  \bibinfo{author}{\bibfnamefont{D.}~\bibnamefont{Yu}},
  \bibinfo{author}{\bibfnamefont{Y.}~\bibnamefont{Singh}},
  \bibinfo{author}{\bibfnamefont{F.}~\bibnamefont{Schreck}},
  \bibinfo{author}{\bibfnamefont{K.}~\bibnamefont{Bongs}}, \bibnamefont{and}
  \bibinfo{author}{\bibfnamefont{I.}~\bibnamefont{Lesanovsky}},
  \bibinfo{journal}{Phys. Rev. Lett.} \textbf{\bibinfo{volume}{110}},
  \bibinfo{pages}{143602} (\bibinfo{year}{2013}).

\bibitem[{\citenamefont{Lahrz et~al.}(2014)\citenamefont{Lahrz, Lemeshko,
  Sengstock, Becker, and Mathey}}]{Lahrz2014dqi}
\bibinfo{author}{\bibfnamefont{M.}~\bibnamefont{Lahrz}},
  \bibinfo{author}{\bibfnamefont{M.}~\bibnamefont{Lemeshko}},
  \bibinfo{author}{\bibfnamefont{K.}~\bibnamefont{Sengstock}},
  \bibinfo{author}{\bibfnamefont{C.}~\bibnamefont{Becker}}, \bibnamefont{and}
  \bibinfo{author}{\bibfnamefont{L.}~\bibnamefont{Mathey}},
  \bibinfo{journal}{arXiv:1402.0873}  (\bibinfo{year}{2014}).

\bibitem[{\citenamefont{Yi et~al.}(2012)\citenamefont{Yi, Diehl, Daley, and
  Zoller}}]{Yi2012ddm}
\bibinfo{author}{\bibfnamefont{W.}~\bibnamefont{Yi}},
  \bibinfo{author}{\bibfnamefont{S.}~\bibnamefont{Diehl}},
  \bibinfo{author}{\bibfnamefont{A.~J.} \bibnamefont{Daley}}, \bibnamefont{and}
  \bibinfo{author}{\bibfnamefont{P.}~\bibnamefont{Zoller}},
  \bibinfo{journal}{New Journal of Physics} \textbf{\bibinfo{volume}{14}},
  \bibinfo{pages}{055002} (\bibinfo{year}{2012}).

\bibitem[{\citenamefont{Boada et~al.}(2012)\citenamefont{Boada, Celi, Latorre,
  and Lewenstein}}]{Boada2012qso}
\bibinfo{author}{\bibfnamefont{O.}~\bibnamefont{Boada}},
  \bibinfo{author}{\bibfnamefont{A.}~\bibnamefont{Celi}},
  \bibinfo{author}{\bibfnamefont{J.~I.} \bibnamefont{Latorre}},
  \bibnamefont{and}
  \bibinfo{author}{\bibfnamefont{M.}~\bibnamefont{Lewenstein}},
  \bibinfo{journal}{Phys. Rev. Lett.} \textbf{\bibinfo{volume}{108}},
  \bibinfo{pages}{133001} (\bibinfo{year}{2012}).

\bibitem[{\citenamefont{Stock et~al.}(2008)\citenamefont{Stock, Babcock,
  Raizen, and Sanders}}]{Stock2008eog}
\bibinfo{author}{\bibfnamefont{R.}~\bibnamefont{Stock}},
  \bibinfo{author}{\bibfnamefont{N.~S.} \bibnamefont{Babcock}},
  \bibinfo{author}{\bibfnamefont{M.~G.} \bibnamefont{Raizen}},
  \bibnamefont{and} \bibinfo{author}{\bibfnamefont{B.~C.}
  \bibnamefont{Sanders}}, \bibinfo{journal}{Phys. Rev. A}
  \textbf{\bibinfo{volume}{78}}, \bibinfo{eid}{022301} (\bibinfo{year}{2008}).

\bibitem[{\citenamefont{Daley et~al.}(2008)\citenamefont{Daley, Boyd, Ye, and
  Zoller}}]{Daley2008qcw}
\bibinfo{author}{\bibfnamefont{A.~J.} \bibnamefont{Daley}},
  \bibinfo{author}{\bibfnamefont{M.~M.} \bibnamefont{Boyd}},
  \bibinfo{author}{\bibfnamefont{J.}~\bibnamefont{Ye}}, \bibnamefont{and}
  \bibinfo{author}{\bibfnamefont{P.}~\bibnamefont{Zoller}},
  \bibinfo{journal}{Phys. Rev. Lett.} \textbf{\bibinfo{volume}{101}},
  \bibinfo{pages}{170504} (\bibinfo{year}{2008}).

\bibitem[{\citenamefont{Gorshkov et~al.}(2009)\citenamefont{Gorshkov, Rey,
  Daley, Boyd, Ye, Zoller, and Lukin}}]{Gorshkov2009aem}
\bibinfo{author}{\bibfnamefont{A.~V.} \bibnamefont{Gorshkov}},
  \bibinfo{author}{\bibfnamefont{A.~M.} \bibnamefont{Rey}},
  \bibinfo{author}{\bibfnamefont{A.~J.} \bibnamefont{Daley}},
  \bibinfo{author}{\bibfnamefont{M.~M.} \bibnamefont{Boyd}},
  \bibinfo{author}{\bibfnamefont{J.}~\bibnamefont{Ye}},
  \bibinfo{author}{\bibfnamefont{P.}~\bibnamefont{Zoller}}, \bibnamefont{and}
  \bibinfo{author}{\bibfnamefont{M.~D.} \bibnamefont{Lukin}},
  \bibinfo{journal}{Phys. Rev. Lett.} \textbf{\bibinfo{volume}{102}},
  \bibinfo{pages}{110503} (\bibinfo{year}{2009}).

\bibitem[{\citenamefont{Reichenbach et~al.}(2009)\citenamefont{Reichenbach,
  Julienne, and Deutsch}}]{Reichenbach2009cns}
\bibinfo{author}{\bibfnamefont{I.}~\bibnamefont{Reichenbach}},
  \bibinfo{author}{\bibfnamefont{P.~S.} \bibnamefont{Julienne}},
  \bibnamefont{and} \bibinfo{author}{\bibfnamefont{I.~H.}
  \bibnamefont{Deutsch}}, \bibinfo{journal}{Phys. Rev. A}
  \textbf{\bibinfo{volume}{80}}, \bibinfo{pages}{020701}
  (\bibinfo{year}{2009}).

\bibitem[{\citenamefont{Daley}(2011)}]{Daley2011qca}
\bibinfo{author}{\bibfnamefont{A.~J.} \bibnamefont{Daley}},
  \bibinfo{journal}{Quantum Inf. Process.} \textbf{\bibinfo{volume}{10}},
  \bibinfo{pages}{865} (\bibinfo{year}{2011}).

\bibitem[{\citenamefont{Rehbein et~al.}(2007)\citenamefont{Rehbein,
  Mehlst\"aubler, Keupp, Moldenhauer, Rasel, Ertmer, Douillet, Michels, Porsev,
  Derevianko et~al.}}]{Rehbein2007oqo}
\bibinfo{author}{\bibfnamefont{N.}~\bibnamefont{Rehbein}},
  \bibinfo{author}{\bibfnamefont{T.~E.} \bibnamefont{Mehlst\"aubler}},
  \bibinfo{author}{\bibfnamefont{J.}~\bibnamefont{Keupp}},
  \bibinfo{author}{\bibfnamefont{K.}~\bibnamefont{Moldenhauer}},
  \bibinfo{author}{\bibfnamefont{E.~M.} \bibnamefont{Rasel}},
  \bibinfo{author}{\bibfnamefont{W.}~\bibnamefont{Ertmer}},
  \bibinfo{author}{\bibfnamefont{A.}~\bibnamefont{Douillet}},
  \bibinfo{author}{\bibfnamefont{V.}~\bibnamefont{Michels}},
  \bibinfo{author}{\bibfnamefont{S.~G.} \bibnamefont{Porsev}},
  \bibinfo{author}{\bibfnamefont{A.}~\bibnamefont{Derevianko}},
  \bibnamefont{et~al.}, \bibinfo{journal}{Phys. Rev. A}
  \textbf{\bibinfo{volume}{76}}, \bibinfo{pages}{043406}
  (\bibinfo{year}{2007}).

\bibitem[{\citenamefont{Binnewies et~al.}(2001)\citenamefont{Binnewies,
  Wilpers, Sterr, Riehle, Helmcke, Mehlst\"aubler, Rasel, and
  Ertmer}}]{Binnewies2001dca}
\bibinfo{author}{\bibfnamefont{T.}~\bibnamefont{Binnewies}},
  \bibinfo{author}{\bibfnamefont{G.}~\bibnamefont{Wilpers}},
  \bibinfo{author}{\bibfnamefont{U.}~\bibnamefont{Sterr}},
  \bibinfo{author}{\bibfnamefont{F.}~\bibnamefont{Riehle}},
  \bibinfo{author}{\bibfnamefont{J.}~\bibnamefont{Helmcke}},
  \bibinfo{author}{\bibfnamefont{T.~E.} \bibnamefont{Mehlst\"aubler}},
  \bibinfo{author}{\bibfnamefont{E.~M.} \bibnamefont{Rasel}}, \bibnamefont{and}
  \bibinfo{author}{\bibfnamefont{W.}~\bibnamefont{Ertmer}},
  \bibinfo{journal}{Phys. Rev. Lett.} \textbf{\bibinfo{volume}{87}},
  \bibinfo{pages}{123002} (\bibinfo{year}{2001}).

\bibitem[{\citenamefont{Katori et~al.}(1999)\citenamefont{Katori, Ido, Isoya,
  and Kuwata-Gonokami}}]{Katori1999mot}
\bibinfo{author}{\bibfnamefont{H.}~\bibnamefont{Katori}},
  \bibinfo{author}{\bibfnamefont{T.}~\bibnamefont{Ido}},
  \bibinfo{author}{\bibfnamefont{Y.}~\bibnamefont{Isoya}}, \bibnamefont{and}
  \bibinfo{author}{\bibfnamefont{M.}~\bibnamefont{Kuwata-Gonokami}},
  \bibinfo{journal}{Phys. Rev. Lett.} \textbf{\bibinfo{volume}{82}},
  \bibinfo{pages}{1116} (\bibinfo{year}{1999}).

\bibitem[{\citenamefont{Gr{\"u}nert and Hemmerich}(2002)}]{Grunert2002sdm}
\bibinfo{author}{\bibfnamefont{J.}~\bibnamefont{Gr{\"u}nert}} \bibnamefont{and}
  \bibinfo{author}{\bibfnamefont{A.}~\bibnamefont{Hemmerich}},
  \bibinfo{journal}{Phys. Rev. A} \textbf{\bibinfo{volume}{65}},
  \bibinfo{pages}{041401} (\bibinfo{year}{2002}).

\bibitem[{\citenamefont{Feldker et~al.}(2011)\citenamefont{Feldker, Sch\"utz,
  John, and Birkl}}]{Feldker2011mot}
\bibinfo{author}{\bibfnamefont{T.}~\bibnamefont{Feldker}},
  \bibinfo{author}{\bibfnamefont{J.}~\bibnamefont{Sch\"utz}},
  \bibinfo{author}{\bibfnamefont{H.}~\bibnamefont{John}}, \bibnamefont{and}
  \bibinfo{author}{\bibfnamefont{G.}~\bibnamefont{Birkl}},
  \bibinfo{journal}{Eur. Phys. J. D} \textbf{\bibinfo{volume}{65}},
  \bibinfo{pages}{257} (\bibinfo{year}{2011}).

\bibitem[{\citenamefont{Porsev et~al.}(2008)\citenamefont{Porsev, Ludlow, Boyd,
  and Ye}}]{Porsev2008dos}
\bibinfo{author}{\bibfnamefont{S.~G.} \bibnamefont{Porsev}},
  \bibinfo{author}{\bibfnamefont{A.~D.} \bibnamefont{Ludlow}},
  \bibinfo{author}{\bibfnamefont{M.~M.} \bibnamefont{Boyd}}, \bibnamefont{and}
  \bibinfo{author}{\bibfnamefont{J.}~\bibnamefont{Ye}}, \bibinfo{journal}{Phys.
  Rev. A} \textbf{\bibinfo{volume}{78}}, \bibinfo{pages}{032508}
  (\bibinfo{year}{2008}).

\bibitem[{\citenamefont{Safronova et~al.}(2013)\citenamefont{Safronova, Porsev,
  Safronova, Kozlov, and Clark}}]{Safranova2013brs}
\bibinfo{author}{\bibfnamefont{M.~S.} \bibnamefont{Safronova}},
  \bibinfo{author}{\bibfnamefont{S.~G.} \bibnamefont{Porsev}},
  \bibinfo{author}{\bibfnamefont{U.~I.} \bibnamefont{Safronova}},
  \bibinfo{author}{\bibfnamefont{M.~G.} \bibnamefont{Kozlov}},
  \bibnamefont{and} \bibinfo{author}{\bibfnamefont{C.~W.} \bibnamefont{Clark}},
  \bibinfo{journal}{Phys. Rev. A} \textbf{\bibinfo{volume}{87}},
  \bibinfo{pages}{012509} (\bibinfo{year}{2013}).

\bibitem[{\citenamefont{Kato et~al.}(2013)\citenamefont{Kato, Sugawa, Shibata,
  Yamamoto, and Takahashi}}]{Kato2013cor}
\bibinfo{author}{\bibfnamefont{S.}~\bibnamefont{Kato}},
  \bibinfo{author}{\bibfnamefont{S.}~\bibnamefont{Sugawa}},
  \bibinfo{author}{\bibfnamefont{K.}~\bibnamefont{Shibata}},
  \bibinfo{author}{\bibfnamefont{R.}~\bibnamefont{Yamamoto}}, \bibnamefont{and}
  \bibinfo{author}{\bibfnamefont{Y.}~\bibnamefont{Takahashi}},
  \bibinfo{journal}{Phys. Rev. Lett.} \textbf{\bibinfo{volume}{110}},
  \bibinfo{pages}{173201} (\bibinfo{year}{2013}).

\bibitem[{\citenamefont{Sansonetti and Nave}(2010)}]{Sansonetti2010wtp}
\bibinfo{author}{\bibfnamefont{J.~E.} \bibnamefont{Sansonetti}}
  \bibnamefont{and} \bibinfo{author}{\bibfnamefont{G.}~\bibnamefont{Nave}},
  \bibinfo{journal}{J. Phys. Chem. Ref. Data} \textbf{\bibinfo{volume}{39}},
  \bibinfo{pages}{033103} (\bibinfo{year}{2010}).

\bibitem[{\citenamefont{Courtillot et~al.}(2005)\citenamefont{Courtillot,
  Quessada-Vial, Brusch, Kolker, Rovera, and Lemonde}}]{Courtillot2005aso}
\bibinfo{author}{\bibfnamefont{I.}~\bibnamefont{Courtillot}},
  \bibinfo{author}{\bibfnamefont{A.}~\bibnamefont{Quessada-Vial}},
  \bibinfo{author}{\bibfnamefont{A.}~\bibnamefont{Brusch}},
  \bibinfo{author}{\bibfnamefont{D.}~\bibnamefont{Kolker}},
  \bibinfo{author}{\bibfnamefont{G.~D.} \bibnamefont{Rovera}},
  \bibnamefont{and} \bibinfo{author}{\bibfnamefont{P.}~\bibnamefont{Lemonde}},
  \bibinfo{journal}{Eur. Phys. J. D} \textbf{\bibinfo{volume}{33}},
  \bibinfo{pages}{161} (\bibinfo{year}{2005}).

\bibitem[{\citenamefont{Beigang et~al.}(1982)\citenamefont{Beigang, L\"ucke,
  Schmidt, Timmermann, and West}}]{Beigang1982opl}
\bibinfo{author}{\bibfnamefont{R.}~\bibnamefont{Beigang}},
  \bibinfo{author}{\bibfnamefont{K.}~\bibnamefont{L\"ucke}},
  \bibinfo{author}{\bibfnamefont{D.}~\bibnamefont{Schmidt}},
  \bibinfo{author}{\bibfnamefont{A.}~\bibnamefont{Timmermann}},
  \bibnamefont{and} \bibinfo{author}{\bibfnamefont{P.~J.} \bibnamefont{West}},
  \bibinfo{journal}{Physica Scripta} \textbf{\bibinfo{volume}{26}},
  \bibinfo{pages}{183} (\bibinfo{year}{1982}).

\bibitem[{\citenamefont{Esherick}(1977)}]{Esherick1977bep}
\bibinfo{author}{\bibfnamefont{P.}~\bibnamefont{Esherick}},
  \bibinfo{journal}{Phys. Rev. A} \textbf{\bibinfo{volume}{15}},
  \bibinfo{pages}{1920} (\bibinfo{year}{1977}).

\bibitem[{\citenamefont{Mickelson et~al.}(2009)\citenamefont{Mickelson,
  Martinez~de Escobar, Anzel, DeSalvo, Nagel, Traverso, Yan, and
  Killian}}]{Mickelson2009ras}
\bibinfo{author}{\bibfnamefont{P.~G.} \bibnamefont{Mickelson}},
  \bibinfo{author}{\bibfnamefont{Y.~N.} \bibnamefont{Martinez~de Escobar}},
  \bibinfo{author}{\bibfnamefont{P.}~\bibnamefont{Anzel}},
  \bibinfo{author}{\bibfnamefont{B.~J.} \bibnamefont{DeSalvo}},
  \bibinfo{author}{\bibfnamefont{S.~B.} \bibnamefont{Nagel}},
  \bibinfo{author}{\bibfnamefont{A.~J.} \bibnamefont{Traverso}},
  \bibinfo{author}{\bibfnamefont{M.}~\bibnamefont{Yan}}, \bibnamefont{and}
  \bibinfo{author}{\bibfnamefont{T.~C.} \bibnamefont{Killian}},
  \bibinfo{journal}{J. Phys. B: At. Mol. Opt. Phys.}
  \textbf{\bibinfo{volume}{42}}, \bibinfo{pages}{235001}
  (\bibinfo{year}{2009}).

\bibitem[{\citenamefont{Dammalapati et~al.}(2011)\citenamefont{Dammalapati,
  Norris, Burrows, and Riis}}]{Dammalapati2011lso}
\bibinfo{author}{\bibfnamefont{U.}~\bibnamefont{Dammalapati}},
  \bibinfo{author}{\bibfnamefont{I.}~\bibnamefont{Norris}},
  \bibinfo{author}{\bibfnamefont{C.}~\bibnamefont{Burrows}}, \bibnamefont{and}
  \bibinfo{author}{\bibfnamefont{E.}~\bibnamefont{Riis}},
  \bibinfo{journal}{Phys. Rev. A} \textbf{\bibinfo{volume}{83}},
  \bibinfo{pages}{062513} (\bibinfo{year}{2011}).

\bibitem[{\citenamefont{Jensen et~al.}(2011)\citenamefont{Jensen, Ming,
  Westergaard, Gunnarsson, Madsen, Brusch, Hald, and Thomsen}}]{Jensen2011edo}
\bibinfo{author}{\bibfnamefont{B.~B.} \bibnamefont{Jensen}},
  \bibinfo{author}{\bibfnamefont{H.}~\bibnamefont{Ming}},
  \bibinfo{author}{\bibfnamefont{P.~G.} \bibnamefont{Westergaard}},
  \bibinfo{author}{\bibfnamefont{K.}~\bibnamefont{Gunnarsson}},
  \bibinfo{author}{\bibfnamefont{M.~H.} \bibnamefont{Madsen}},
  \bibinfo{author}{\bibfnamefont{A.}~\bibnamefont{Brusch}},
  \bibinfo{author}{\bibfnamefont{J.}~\bibnamefont{Hald}}, \bibnamefont{and}
  \bibinfo{author}{\bibfnamefont{J.~W.} \bibnamefont{Thomsen}},
  \bibinfo{journal}{Phys. Rev. Lett.} \textbf{\bibinfo{volume}{107}},
  \bibinfo{pages}{113001} (\bibinfo{year}{2011}).

\bibitem[{\citenamefont{Beloy et~al.}(2012)\citenamefont{Beloy, Sherman, Lemke,
  Hinkley, Oates, and Ludlow}}]{Beloy2012dot}
\bibinfo{author}{\bibfnamefont{K.}~\bibnamefont{Beloy}},
  \bibinfo{author}{\bibfnamefont{J.~A.} \bibnamefont{Sherman}},
  \bibinfo{author}{\bibfnamefont{N.~D.} \bibnamefont{Lemke}},
  \bibinfo{author}{\bibfnamefont{N.}~\bibnamefont{Hinkley}},
  \bibinfo{author}{\bibfnamefont{C.~W.} \bibnamefont{Oates}}, \bibnamefont{and}
  \bibinfo{author}{\bibfnamefont{A.~D.} \bibnamefont{Ludlow}},
  \bibinfo{journal}{Phys. Rev. A} \textbf{\bibinfo{volume}{86}},
  \bibinfo{pages}{051404} (\bibinfo{year}{2012}).

\bibitem[{\citenamefont{McClelland and Hanssen}(2006)}]{McClelland2006lcw}
\bibinfo{author}{\bibfnamefont{J.~J.} \bibnamefont{McClelland}}
  \bibnamefont{and} \bibinfo{author}{\bibfnamefont{J.~L.}
  \bibnamefont{Hanssen}}, \bibinfo{journal}{Phys. Rev. Lett.}
  \textbf{\bibinfo{volume}{96}}, \bibinfo{pages}{143005}
  (\bibinfo{year}{2006}).

\bibitem[{\citenamefont{Lu et~al.}(2010)\citenamefont{Lu, Youn, and
  Lev}}]{Lu2010tud}
\bibinfo{author}{\bibfnamefont{M.}~\bibnamefont{Lu}},
  \bibinfo{author}{\bibfnamefont{S.~H.} \bibnamefont{Youn}}, \bibnamefont{and}
  \bibinfo{author}{\bibfnamefont{B.~L.} \bibnamefont{Lev}},
  \bibinfo{journal}{Phys. Rev. Lett.} \textbf{\bibinfo{volume}{104}},
  \bibinfo{pages}{063001} (\bibinfo{year}{2010}).

\bibitem[{\citenamefont{Bidel}(2002)}]{Bidel2002per}
\bibinfo{author}{\bibfnamefont{Y.}~\bibnamefont{Bidel}}, Ph.D. thesis,
  \bibinfo{school}{Universit\'e de Nice} (\bibinfo{year}{2002}).

\bibitem[{\citenamefont{Porsev et~al.}(1999)\citenamefont{Porsev, Rakhlina, and
  Kozlov}}]{Porsev1999eda}
\bibinfo{author}{\bibfnamefont{S.~G.} \bibnamefont{Porsev}},
  \bibinfo{author}{\bibfnamefont{Y.~G.} \bibnamefont{Rakhlina}},
  \bibnamefont{and} \bibinfo{author}{\bibfnamefont{M.~G.}
  \bibnamefont{Kozlov}}, \bibinfo{journal}{Phys. Rev. A}
  \textbf{\bibinfo{volume}{60}}, \bibinfo{pages}{2781} (\bibinfo{year}{1999}).

\bibitem[{\citenamefont{Cho et~al.}(2012)\citenamefont{Cho, Lee, Lee, Ahn, Lee,
  Yu, Lee, and Park}}]{Cho2012oro}
\bibinfo{author}{\bibfnamefont{J.~W.} \bibnamefont{Cho}},
  \bibinfo{author}{\bibfnamefont{H.-G.} \bibnamefont{Lee}},
  \bibinfo{author}{\bibfnamefont{S.}~\bibnamefont{Lee}},
  \bibinfo{author}{\bibfnamefont{J.}~\bibnamefont{Ahn}},
  \bibinfo{author}{\bibfnamefont{W.-K.} \bibnamefont{Lee}},
  \bibinfo{author}{\bibfnamefont{D.-H.} \bibnamefont{Yu}},
  \bibinfo{author}{\bibfnamefont{S.~K.} \bibnamefont{Lee}}, \bibnamefont{and}
  \bibinfo{author}{\bibfnamefont{C.~Y.} \bibnamefont{Park}},
  \bibinfo{journal}{Phys. Rev. A} \textbf{\bibinfo{volume}{85}},
  \bibinfo{pages}{035401} (\bibinfo{year}{2012}).

\bibitem[{\citenamefont{Santra et~al.}(2004)\citenamefont{Santra, Christ, and
  Greene}}]{Santra2004pom}
\bibinfo{author}{\bibfnamefont{R.}~\bibnamefont{Santra}},
  \bibinfo{author}{\bibfnamefont{K.~V.} \bibnamefont{Christ}},
  \bibnamefont{and} \bibinfo{author}{\bibfnamefont{C.~H.}
  \bibnamefont{Greene}}, \bibinfo{journal}{Phys. Rev. A}
  \textbf{\bibinfo{volume}{69}}, \bibinfo{pages}{042510}
  (\bibinfo{year}{2004}).

\bibitem[{\citenamefont{Yasuda and Katori}(2004)}]{Yasuda2004lmo}
\bibinfo{author}{\bibfnamefont{M.}~\bibnamefont{Yasuda}} \bibnamefont{and}
  \bibinfo{author}{\bibfnamefont{H.}~\bibnamefont{Katori}},
  \bibinfo{journal}{Phys. Rev. Lett.} \textbf{\bibinfo{volume}{92}},
  \bibinfo{pages}{153004} (\bibinfo{year}{2004}).

\bibitem[{\citenamefont{Xu et~al.}(2003)\citenamefont{Xu, Loftus, Hall,
  Gallagher, and Ye}}]{Xu2003cat}
\bibinfo{author}{\bibfnamefont{X.}~\bibnamefont{Xu}},
  \bibinfo{author}{\bibfnamefont{T.~H.} \bibnamefont{Loftus}},
  \bibinfo{author}{\bibfnamefont{J.~L.} \bibnamefont{Hall}},
  \bibinfo{author}{\bibfnamefont{A.}~\bibnamefont{Gallagher}},
  \bibnamefont{and} \bibinfo{author}{\bibfnamefont{J.}~\bibnamefont{Ye}},
  \bibinfo{journal}{J. Opt. Soc. Am. B} \textbf{\bibinfo{volume}{20}},
  \bibinfo{pages}{968} (\bibinfo{year}{2003}).

\bibitem[{\citenamefont{Katori et~al.}(2001)\citenamefont{Katori, Ido, Isoya,
  and Kuwata-Gonokami}}]{Katori2001lco}
\bibinfo{author}{\bibfnamefont{H.}~\bibnamefont{Katori}},
  \bibinfo{author}{\bibfnamefont{T.}~\bibnamefont{Ido}},
  \bibinfo{author}{\bibfnamefont{Y.}~\bibnamefont{Isoya}}, \bibnamefont{and}
  \bibinfo{author}{\bibfnamefont{M.}~\bibnamefont{Kuwata-Gonokami}}, in
  \emph{\bibinfo{booktitle}{Atomic Physics 17}}, edited by
  \bibinfo{editor}{\bibfnamefont{E.}~\bibnamefont{Arimondo}},
  \bibinfo{editor}{\bibfnamefont{P.}~\bibnamefont{DeNatale}}, \bibnamefont{and}
  \bibinfo{editor}{\bibfnamefont{M.}~\bibnamefont{Inguscio}}
  (\bibinfo{publisher}{American Institute of Physics, Woodbury},
  \bibinfo{year}{2001}), pp. \bibinfo{pages}{382--396}.

\bibitem[{\citenamefont{Kurosu and Shimizu}(1992)}]{Kurosu1992lca}
\bibinfo{author}{\bibfnamefont{T.}~\bibnamefont{Kurosu}} \bibnamefont{and}
  \bibinfo{author}{\bibfnamefont{F.}~\bibnamefont{Shimizu}},
  \bibinfo{journal}{Jpn. J. Appl. Phys.} \textbf{\bibinfo{volume}{31}},
  \bibinfo{pages}{908} (\bibinfo{year}{1992}).

\bibitem[{\citenamefont{Dinneen et~al.}(1999)\citenamefont{Dinneen, Vogel,
  Arimondo, Hall, and Gallagher}}]{Dinneen1999cco}
\bibinfo{author}{\bibfnamefont{T.~P.} \bibnamefont{Dinneen}},
  \bibinfo{author}{\bibfnamefont{K.~R.} \bibnamefont{Vogel}},
  \bibinfo{author}{\bibfnamefont{E.}~\bibnamefont{Arimondo}},
  \bibinfo{author}{\bibfnamefont{J.~L.} \bibnamefont{Hall}}, \bibnamefont{and}
  \bibinfo{author}{\bibfnamefont{A.}~\bibnamefont{Gallagher}},
  \bibinfo{journal}{Phys. Rev. A} \textbf{\bibinfo{volume}{59}},
  \bibinfo{pages}{1216} (\bibinfo{year}{1999}).

\bibitem[{\citenamefont{Poli et~al.}(2005)\citenamefont{Poli, Drullinger,
  Ferrari, L\'eonard, Sorrentino, and Tino}}]{Poli2005cat}
\bibinfo{author}{\bibfnamefont{N.}~\bibnamefont{Poli}},
  \bibinfo{author}{\bibfnamefont{R.~E.} \bibnamefont{Drullinger}},
  \bibinfo{author}{\bibfnamefont{G.}~\bibnamefont{Ferrari}},
  \bibinfo{author}{\bibfnamefont{J.}~\bibnamefont{L\'eonard}},
  \bibinfo{author}{\bibfnamefont{F.}~\bibnamefont{Sorrentino}},
  \bibnamefont{and} \bibinfo{author}{\bibfnamefont{G.~M.} \bibnamefont{Tino}},
  \bibinfo{journal}{Phys. Rev. A} \textbf{\bibinfo{volume}{71}},
  \bibinfo{pages}{061403} (\bibinfo{year}{2005}).

\bibitem[{Kil()}]{KillianPrivComm}
\bibinfo{note}{T.~C.~Killian (Rice University), private communication.}

\bibitem[{\citenamefont{Stuhler et~al.}(2001)\citenamefont{Stuhler, Schmidt,
  Hensler, Werner, Mlynek, and Pfau}}]{Stuhler2001clo}
\bibinfo{author}{\bibfnamefont{J.}~\bibnamefont{Stuhler}},
  \bibinfo{author}{\bibfnamefont{P.~O.} \bibnamefont{Schmidt}},
  \bibinfo{author}{\bibfnamefont{S.}~\bibnamefont{Hensler}},
  \bibinfo{author}{\bibfnamefont{J.}~\bibnamefont{Werner}},
  \bibinfo{author}{\bibfnamefont{J.}~\bibnamefont{Mlynek}}, \bibnamefont{and}
  \bibinfo{author}{\bibfnamefont{T.}~\bibnamefont{Pfau}},
  \bibinfo{journal}{Phys. Rev. A} \textbf{\bibinfo{volume}{64}},
  \bibinfo{pages}{031405(R)} (\bibinfo{year}{2001}).

\bibitem[{\citenamefont{Traverso et~al.}(2009)\citenamefont{Traverso,
  Chakraborty, {Martinez de Escobar}, Mickelson, Nagel, Yan, and
  Killian}}]{Traverso2009iae}
\bibinfo{author}{\bibfnamefont{A.}~\bibnamefont{Traverso}},
  \bibinfo{author}{\bibfnamefont{R.}~\bibnamefont{Chakraborty}},
  \bibinfo{author}{\bibfnamefont{Y.~N.} \bibnamefont{{Martinez de Escobar}}},
  \bibinfo{author}{\bibfnamefont{P.~G.} \bibnamefont{Mickelson}},
  \bibinfo{author}{\bibfnamefont{S.~B.} \bibnamefont{Nagel}},
  \bibinfo{author}{\bibfnamefont{M.}~\bibnamefont{Yan}}, \bibnamefont{and}
  \bibinfo{author}{\bibfnamefont{T.~C.} \bibnamefont{Killian}},
  \bibinfo{journal}{Phys. Rev. A} \textbf{\bibinfo{volume}{79}},
  \bibinfo{pages}{060702(R)} (\bibinfo{year}{2009}).

\bibitem[{\citenamefont{Mukaiyama et~al.}(2003)\citenamefont{Mukaiyama, Katori,
  Ido, Li, and Kuwata-Gonokami}}]{Mukaiyama2003rll}
\bibinfo{author}{\bibfnamefont{T.}~\bibnamefont{Mukaiyama}},
  \bibinfo{author}{\bibfnamefont{H.}~\bibnamefont{Katori}},
  \bibinfo{author}{\bibfnamefont{T.}~\bibnamefont{Ido}},
  \bibinfo{author}{\bibfnamefont{Y.}~\bibnamefont{Li}}, \bibnamefont{and}
  \bibinfo{author}{\bibfnamefont{M.}~\bibnamefont{Kuwata-Gonokami}},
  \bibinfo{journal}{Phys. Rev. Lett.} \textbf{\bibinfo{volume}{90}},
  \bibinfo{pages}{113002} (\bibinfo{year}{2003}).

\bibitem[{\citenamefont{Heider and Brink}(1977)}]{Heider1977hso}
\bibinfo{author}{\bibfnamefont{S.~M.} \bibnamefont{Heider}} \bibnamefont{and}
  \bibinfo{author}{\bibfnamefont{G.~O.} \bibnamefont{Brink}},
  \bibinfo{journal}{Phys. Rev. A} \textbf{\bibinfo{volume}{16}},
  \bibinfo{pages}{1371} (\bibinfo{year}{1977}).

\bibitem[{\citenamefont{Bushaw et~al.}(1993)\citenamefont{Bushaw, Kluge,
  Lantzsch, Schwalbach, Stenner, Stevens, Wendt, and Zimmer}}]{Bushaw1993hsi}
\bibinfo{author}{\bibfnamefont{B.}~\bibnamefont{Bushaw}},
  \bibinfo{author}{\bibfnamefont{H.-J.} \bibnamefont{Kluge}},
  \bibinfo{author}{\bibfnamefont{J.}~\bibnamefont{Lantzsch}},
  \bibinfo{author}{\bibfnamefont{R.}~\bibnamefont{Schwalbach}},
  \bibinfo{author}{\bibfnamefont{J.}~\bibnamefont{Stenner}},
  \bibinfo{author}{\bibfnamefont{H.}~\bibnamefont{Stevens}},
  \bibinfo{author}{\bibfnamefont{K.}~\bibnamefont{Wendt}}, \bibnamefont{and}
  \bibinfo{author}{\bibfnamefont{K.}~\bibnamefont{Zimmer}},
  \bibinfo{journal}{Zeitschrift f\"ur Physik D} \textbf{\bibinfo{volume}{28}},
  \bibinfo{pages}{275} (\bibinfo{year}{1993}).

\bibitem[{\citenamefont{Werij et~al.}(1992)\citenamefont{Werij, Greene,
  Theodosiou, and Gallagher}}]{Werij1992osa}
\bibinfo{author}{\bibfnamefont{H.~G.~C.} \bibnamefont{Werij}},
  \bibinfo{author}{\bibfnamefont{C.~H.} \bibnamefont{Greene}},
  \bibinfo{author}{\bibfnamefont{C.~E.} \bibnamefont{Theodosiou}},
  \bibnamefont{and}
  \bibinfo{author}{\bibfnamefont{A.}~\bibnamefont{Gallagher}},
  \bibinfo{journal}{Phys. Rev. A} \textbf{\bibinfo{volume}{46}},
  \bibinfo{pages}{1248} (\bibinfo{year}{1992}).

\bibitem[{\citenamefont{Ludlow}(2008)}]{Ludlow2008tso}
\bibinfo{author}{\bibfnamefont{A.~D.} \bibnamefont{Ludlow}}, Ph.D. thesis,
  \bibinfo{school}{University of Colorado, Boulder, USA}
  (\bibinfo{year}{2008}).

\bibitem[{\citenamefont{Sorrentino et~al.}(2006)\citenamefont{Sorrentino,
  Ferrari, Poli, Drullinger, and Tino}}]{Sorrentino2006lca}
\bibinfo{author}{\bibfnamefont{F.}~\bibnamefont{Sorrentino}},
  \bibinfo{author}{\bibfnamefont{G.}~\bibnamefont{Ferrari}},
  \bibinfo{author}{\bibfnamefont{N.}~\bibnamefont{Poli}},
  \bibinfo{author}{\bibfnamefont{R.}~\bibnamefont{Drullinger}},
  \bibnamefont{and} \bibinfo{author}{\bibfnamefont{G.~M.} \bibnamefont{Tino}},
  \bibinfo{journal}{Mod. Phys. Lett. B} \textbf{\bibinfo{volume}{20}},
  \bibinfo{pages}{1287} (\bibinfo{year}{2006}).

\bibitem[{Mar()}]{MartinPrivComm}
\bibinfo{note}{M.~J.~Martin (JILA), private communication.}

\bibitem[{\citenamefont{Guest et~al.}(2007)\citenamefont{Guest, Scielzo, Ahmad,
  Bailey, Greene, Holt, Lu, O'Connor, and Potterveld}}]{Guest2007lto}
\bibinfo{author}{\bibfnamefont{J.~R.} \bibnamefont{Guest}},
  \bibinfo{author}{\bibfnamefont{N.~D.} \bibnamefont{Scielzo}},
  \bibinfo{author}{\bibfnamefont{I.}~\bibnamefont{Ahmad}},
  \bibinfo{author}{\bibfnamefont{K.}~\bibnamefont{Bailey}},
  \bibinfo{author}{\bibfnamefont{J.~P.} \bibnamefont{Greene}},
  \bibinfo{author}{\bibfnamefont{R.~J.} \bibnamefont{Holt}},
  \bibinfo{author}{\bibfnamefont{Z.-T.} \bibnamefont{Lu}},
  \bibinfo{author}{\bibfnamefont{T.~P.} \bibnamefont{O'Connor}},
  \bibnamefont{and} \bibinfo{author}{\bibfnamefont{D.~H.}
  \bibnamefont{Potterveld}}, \bibinfo{journal}{Phys. Rev. Lett.}
  \textbf{\bibinfo{volume}{98}}, \bibinfo{pages}{093001}
  (\bibinfo{year}{2007}).

\bibitem[{\citenamefont{Hansen and Hemmerich}(2005)}]{Hansen2005dfs}
\bibinfo{author}{\bibfnamefont{D.}~\bibnamefont{Hansen}} \bibnamefont{and}
  \bibinfo{author}{\bibfnamefont{A.}~\bibnamefont{Hemmerich}},
  \bibinfo{journal}{Phys. Rev. A} \textbf{\bibinfo{volume}{72}},
  \bibinfo{pages}{022502} (\bibinfo{year}{2005}).

\bibitem[{\citenamefont{Hancox et~al.}(2004)\citenamefont{Hancox, Doret,
  Hummon, Luo, and Doyle}}]{Hancox2004mto}
\bibinfo{author}{\bibfnamefont{C.~I.} \bibnamefont{Hancox}},
  \bibinfo{author}{\bibfnamefont{S.~C.} \bibnamefont{Doret}},
  \bibinfo{author}{\bibfnamefont{M.~T.} \bibnamefont{Hummon}},
  \bibinfo{author}{\bibfnamefont{L.}~\bibnamefont{Luo}}, \bibnamefont{and}
  \bibinfo{author}{\bibfnamefont{J.~M.} \bibnamefont{Doyle}},
  \bibinfo{journal}{Nature} \textbf{\bibinfo{volume}{431}},
  \bibinfo{pages}{281} (\bibinfo{year}{2004}).

\bibitem[{\citenamefont{Lu et~al.}(2011)\citenamefont{Lu, Burdick, Youn, and
  Lev}}]{Lu2011sdb}
\bibinfo{author}{\bibfnamefont{M.}~\bibnamefont{Lu}},
  \bibinfo{author}{\bibfnamefont{N.~Q.} \bibnamefont{Burdick}},
  \bibinfo{author}{\bibfnamefont{S.~H.} \bibnamefont{Youn}}, \bibnamefont{and}
  \bibinfo{author}{\bibfnamefont{B.~L.} \bibnamefont{Lev}},
  \bibinfo{journal}{Phys. Rev. Lett.} \textbf{\bibinfo{volume}{107}},
  \bibinfo{pages}{190401} (\bibinfo{year}{2011}).

\bibitem[{\citenamefont{Miao et~al.}(2014)\citenamefont{Miao, Hostetter,
  Stratis, and Saffman}}]{Miao2014mot}
\bibinfo{author}{\bibfnamefont{J.}~\bibnamefont{Miao}},
  \bibinfo{author}{\bibfnamefont{J.}~\bibnamefont{Hostetter}},
  \bibinfo{author}{\bibfnamefont{G.}~\bibnamefont{Stratis}}, \bibnamefont{and}
  \bibinfo{author}{\bibfnamefont{M.}~\bibnamefont{Saffman}},
  \bibinfo{journal}{Phys. Rev. A} \textbf{\bibinfo{volume}{89}},
  \bibinfo{pages}{041401} (\bibinfo{year}{2014}).

\bibitem[{\citenamefont{Aikawa et~al.}(2012)\citenamefont{Aikawa, Frisch, Mark,
  Baier, Rietzler, Grimm, and Ferlaino}}]{Aikawa2012bec}
\bibinfo{author}{\bibfnamefont{K.}~\bibnamefont{Aikawa}},
  \bibinfo{author}{\bibfnamefont{A.}~\bibnamefont{Frisch}},
  \bibinfo{author}{\bibfnamefont{M.}~\bibnamefont{Mark}},
  \bibinfo{author}{\bibfnamefont{S.}~\bibnamefont{Baier}},
  \bibinfo{author}{\bibfnamefont{A.}~\bibnamefont{Rietzler}},
  \bibinfo{author}{\bibfnamefont{R.}~\bibnamefont{Grimm}}, \bibnamefont{and}
  \bibinfo{author}{\bibfnamefont{F.}~\bibnamefont{Ferlaino}},
  \bibinfo{journal}{Phys. Rev. Lett.} \textbf{\bibinfo{volume}{108}},
  \bibinfo{pages}{210401} (\bibinfo{year}{2012}).

\bibitem[{\citenamefont{Sukachev et~al.}(2010)\citenamefont{Sukachev, Sokolov,
  Chebakov, Akimov, Kanorsky, Kolachevsky, and Sorokin}}]{Sukachev2010mot}
\bibinfo{author}{\bibfnamefont{D.}~\bibnamefont{Sukachev}},
  \bibinfo{author}{\bibfnamefont{A.}~\bibnamefont{Sokolov}},
  \bibinfo{author}{\bibfnamefont{K.}~\bibnamefont{Chebakov}},
  \bibinfo{author}{\bibfnamefont{A.}~\bibnamefont{Akimov}},
  \bibinfo{author}{\bibfnamefont{S.}~\bibnamefont{Kanorsky}},
  \bibinfo{author}{\bibfnamefont{N.}~\bibnamefont{Kolachevsky}},
  \bibnamefont{and} \bibinfo{author}{\bibfnamefont{V.}~\bibnamefont{Sorokin}},
  \bibinfo{journal}{Phys. Rev. A} \textbf{\bibinfo{volume}{82}},
  \bibinfo{pages}{011405} (\bibinfo{year}{2010}).

\bibitem[{\citenamefont{Chwalla et~al.}(2009)\citenamefont{Chwalla, Benhelm,
  Kim, Kirchmair, Monz, Riebe, Schindler, Villar, H\"ansel, Roos
  et~al.}}]{Chwalla2009afm}
\bibinfo{author}{\bibfnamefont{M.}~\bibnamefont{Chwalla}},
  \bibinfo{author}{\bibfnamefont{J.}~\bibnamefont{Benhelm}},
  \bibinfo{author}{\bibfnamefont{K.}~\bibnamefont{Kim}},
  \bibinfo{author}{\bibfnamefont{G.}~\bibnamefont{Kirchmair}},
  \bibinfo{author}{\bibfnamefont{T.}~\bibnamefont{Monz}},
  \bibinfo{author}{\bibfnamefont{M.}~\bibnamefont{Riebe}},
  \bibinfo{author}{\bibfnamefont{P.}~\bibnamefont{Schindler}},
  \bibinfo{author}{\bibfnamefont{A.~S.} \bibnamefont{Villar}},
  \bibinfo{author}{\bibfnamefont{W.}~\bibnamefont{H\"ansel}},
  \bibinfo{author}{\bibfnamefont{C.~F.} \bibnamefont{Roos}},
  \bibnamefont{et~al.}, \bibinfo{journal}{Phys. Rev. Lett.}
  \textbf{\bibinfo{volume}{102}}, \bibinfo{pages}{023002}
  (\bibinfo{year}{2009}).

\bibitem[{\citenamefont{Wolf et~al.}(2008)\citenamefont{Wolf, van~den Berg,
  Gohle, Salumbides, Ubachs, and Eikema}}]{Wolf2008fmo}
\bibinfo{author}{\bibfnamefont{A.~L.} \bibnamefont{Wolf}},
  \bibinfo{author}{\bibfnamefont{S.~A.} \bibnamefont{van~den Berg}},
  \bibinfo{author}{\bibfnamefont{C.}~\bibnamefont{Gohle}},
  \bibinfo{author}{\bibfnamefont{E.~J.} \bibnamefont{Salumbides}},
  \bibinfo{author}{\bibfnamefont{W.}~\bibnamefont{Ubachs}}, \bibnamefont{and}
  \bibinfo{author}{\bibfnamefont{K.~S.~E.} \bibnamefont{Eikema}},
  \bibinfo{journal}{Phys. Rev. A} \textbf{\bibinfo{volume}{78}},
  \bibinfo{pages}{032511} (\bibinfo{year}{2008}).

\bibitem[{\citenamefont{Ferrari et~al.}(2003)\citenamefont{Ferrari, Cancio,
  Drullinger, Giusfredi, Poli, Prevedelli, Toninelli, and
  Tino}}]{Ferrari2003pfm}
\bibinfo{author}{\bibfnamefont{G.}~\bibnamefont{Ferrari}},
  \bibinfo{author}{\bibfnamefont{P.}~\bibnamefont{Cancio}},
  \bibinfo{author}{\bibfnamefont{R.}~\bibnamefont{Drullinger}},
  \bibinfo{author}{\bibfnamefont{G.}~\bibnamefont{Giusfredi}},
  \bibinfo{author}{\bibfnamefont{N.}~\bibnamefont{Poli}},
  \bibinfo{author}{\bibfnamefont{M.}~\bibnamefont{Prevedelli}},
  \bibinfo{author}{\bibfnamefont{C.}~\bibnamefont{Toninelli}},
  \bibnamefont{and} \bibinfo{author}{\bibfnamefont{G.~M.} \bibnamefont{Tino}},
  \bibinfo{journal}{Phys. Rev. Lett.} \textbf{\bibinfo{volume}{91}},
  \bibinfo{pages}{243002} (\bibinfo{year}{2003}).

\bibitem[{\citenamefont{Ye et~al.}(1996)\citenamefont{Ye, Swartz, Jungner, and
  Hall}}]{Ye1996hsa}
\bibinfo{author}{\bibfnamefont{J.}~\bibnamefont{Ye}},
  \bibinfo{author}{\bibfnamefont{S.}~\bibnamefont{Swartz}},
  \bibinfo{author}{\bibfnamefont{P.}~\bibnamefont{Jungner}}, \bibnamefont{and}
  \bibinfo{author}{\bibfnamefont{J.~L.} \bibnamefont{Hall}},
  \bibinfo{journal}{Opt. Lett.} \textbf{\bibinfo{volume}{21}},
  \bibinfo{pages}{1280} (\bibinfo{year}{1996}).

\end{thebibliography}

%\end{document}

\end{document}